\documentclass[]{aastex}
\pdfoutput=1
\usepackage{graphicx}
\usepackage{natbib}
\bibpunct{(}{)}{;}{a}{}{,} 
\usepackage{amsmath}
\usepackage{amssymb}
\usepackage{wasysym}
\usepackage{ifpdf}
\usepackage{url}

\begin{document}

\title{\textit{Aeolus}: A Markov--Chain Monte Carlo code for mapping ultracool atmospheres. \\
An application on Jupiter and brown dwarf HST light curves.  }

\author{Theodora Karalidi}
\affil{Steward Observatory, Department of Astronomy, The University of Arizona, 933 N. Cherry Ave, Tucson, AZ 85721, USA, \\ \textit{tkaralidi@email.arizona.edu}}

\author{D\'aniel Apai}
\affil{Steward Observatory, Department of Astronomy, The University of Arizona, 933 N. Cherry Ave, Tucson, AZ 85721, USA \\
Lunar and Planetary Laboratory, University of Arizona,1629 E University Blvd, AZ 85721, USA \\ 
Earths in Other Solar Systems Team}
\author{Glenn Schneider}
\affil{Steward Observatory, Department of Astronomy, The University of Arizona, 933 N. Cherry Ave, Tucson, AZ 85721, USA}
\author{Jake R. Hanson}
\affil{Steward Observatory, Department of Astronomy, The University of Arizona, 933 N. Cherry Ave, Tucson, AZ 85721, USA}
\and
\author{Jay M. Pasachoff}
\affil{Hopkins Observatory, Williams College, 33 Lab Campus Drive, Williamstown, MA 01267, USA \\
Planetary Sciences Department, Caltech, Pasadena, CA 91125, USA}


\begin{abstract}
Deducing the cloud cover and its temporal evolution from the observed planetary spectra 
and phase curves can give us major insight into the atmospheric dynamics. 
In this paper, we present \textit{Aeolus}, a Markov--Chain Monte Carlo code that maps the 
structure of brown dwarf and other ultracool atmospheres. We validated \textit{Aeolus} on 
a set of \emph{unique} Jupiter Hubble Space Telescope (HST) light curves. \textit{Aeolus} 
accurately retrieves the properties of the major features of the jovian atmosphere such as 
the Great Red Spot and a major 5$\mu$m hot spot. \textit{Aeolus} is  
the first mapping code validated on actual observations of a giant planet over a full 
rotational period. For this study, we applied \textit{Aeolus} to J and H--bands HST light 
curves of 2MASSJ21392676+0220226 and 2MASSJ0136565+093347. \textit{Aeolus} 
retrieves three spots at the top--of--the--atmosphere (per observational wavelength) of 
these two brown dwarfs, with a surface coverage of 21\%$\pm$3\% and 20.3\%$\pm$1.5\% 
respectively. The Jupiter HST light curves will be publicly available via ADS/VIZIR. 
\end{abstract}



\section{Introduction}

High-quality observations of giant exoplanets suggest that their atmospheres at 
high altitudes are dominated by clouds and hazes [i.e., WASP 12b \cite[see, e.g.,][]{sing13}, 
Kepler--7b \citep[e.g.,][]{demory13}, HD 189733 b \citep[e.g.,][]{pont08}, GJ1214b 
\citep[e.g.,][]{bean10,kreidberg14}, and HD 97658b \citep{knutson14}]. 
Similarly, the combination of clouds and vigorous atmospheric dynamics results in time-evolving 
atmospheric features in Solar System giant planets. Episodic bright spots 
have, for example, been observed in Saturn's atmosphere, 
lasting over a year, perturbing the cloud structure of the planet and 
increasing the planetary albedo \citep[][]{west09}; further, 
Neptune and Uranus exhibit episodic dark and/or 
bright spots \citep[][]{sromovsky02,sromovsky12} 
and high zonal wind speeds \citep[][]{irwin11,sromovsky12}.

Radiative transfer models of brown dwarf atmospheres predicted the existence of 
complex cloud structures that lead to time--varying disk--integrated fluxes due to rotational 
modulations \citep[see, e.g.,][]{marley10,morley14}. These predictions were confirmed by 
recent time--resolved observations of L/T and late-T--type brown dwarfs 
\citep[see, e.g.,][]{artigau09, radigan12, apai13, biller13}. 
Models of atmospheric dynamics in brown dwarfs predicts that the vigorous circulation 
and winds will re-arrange the cloud cover on rapid timescales \citep[e.g.][]{showman13, zhang14}.
Consistent with this general prediction light curve evolution has been observed in two brown dwarfs  
observed over more than a single rotational period \citep[][]{artigau09,apai13,buenzli15}.

Hazes are also common in the atmospheres of Solar System 
planets and brown dwarfs. Saturn's and Jupiter's poles 
are covered by a thick layer of stratospheric 
hazes, while the central disk (low latitudes) is covered by clouds and 
hazes rotating at high zonal speeds \citep[][]{west09}. 
Observations of brown dwarfs indicate the existence of hazes 
at high altitudes across the disk \citep[see, e.g.,][]{yang15}.
Even though brown dwarfs usually lack a parent star, and thus don't receive 
UV radiation, hazes could be created by auroral phenomena \citep[][]{pryor91}. 

Atmospheric dynamics, clouds, and hazes have complicated and intertwined roles in ultracool 
atmospheres affecting radiation transport, atmospheric chemistry and 
influencing surface temperatures and potential habitability \citep[][]{marley13}. 
Due to the high complexity of ultracool atmospheres, the study of atmospheric dynamics 
and cloud characterization is difficult. 
A major insight is gained into the atmospheric dynamics when the cloud cover and its 
temporal evolution can be deduced from the observed planetary spectra and phase curves.

To date, a number of exoplanets and 
brown dwarfs have been mapped using various techniques. 
\citet[][]{knutson07}, \citet[][]{dewit12} and \citet[][]{snellen09} 
have used exoplanetary phase curves in combination with homogeneous 
brightness--slice models  and a Markov--Chain Monte Carlo code to acquire 
information on the planetary orbit parameters, as well as possible heterogeneities 
on the planet, and create the surface brightness maps of HD189733b and CoRoT-1b. 
\citet[][]{cowan08} and \citet[][]{cowan13} used planetary 
phase curves with a brightness--slice model and Fourier inversion techniques 
to map modeled exoplanets. These techniques are based on knowing the 
rotation rate of the planet (for these hot Jupiters it is probably equal to their orbital rate) 
and assuming that atmospheric patterns are stable during a full rotational period.
 \citet[][]{apai13} used time-resolved HST spectra to map the brown 
dwarfs 2MASSJ21392676+0220226 (2M2139) and 2MASSJ0136565+093347 (SIMP0136).
In this study they first applied a principal components analysis (PCA) on the spectral cube
 to determine the smallest number of independent spectral components present in the photosphere.
 Then with a Genetic Algorithm--optimized ray tracing model (\textit{Stratos}) they
 identified the simplest models that are consistent with the observed light curve shapes.
 Finally, \citet[][]{crossfield14} used Doppler Imaging to map the nearest--known variable 
 brown dwarf Luhman 16B \citep[][]{luhman13}. Doppler imaging uses measurements of 
rotationally broadened absorption line profiles, and their variations due to 
atmospheric heterogeneities, to map the planetary atmosphere.

Here we present \textit{Aeolus}, a Markov--Chain Monte Carlo (MCMC) code that maps 
the top--of--the--atmosphere structure of brown dwarf and other ultracool atmospheres. 
Because of the use of bayesian inference, an 
MCMC code can fit input observations with high--dimensional models 
(such as the structure of an atmosphere) and can provide more 
precise estimates of uncertainties and correlations in model parameters than other 
commonly used methods. 
Although our code was initially developed to map brown dwarf 
atmospheres, in the future it can be applied to any directly detected (exo)planet 
atmosphere. For example, to validate our code, 
we applied it to HST Jupiter light curves.

As a spatially resolved source, with a wealth of information existing about its 
atmospheric composition and dynamical structure 
\citep[see, e.g.,][]{bagenal04,depaterlissauer10}, Jupiter offers a unique target for 
testing mapping techniques. Jupiter's (latitudinally dependent) rotational period, 
9$^\mathrm{hrs}$55$^\mathrm{m}$27$^\mathrm{s}$.3 
\citet[][]{depaterlissauer10}, is comparable to that of brown dwarfs; Jupiter has a wealth of atmospheric 
features (e.g., Great Red Spot, hot spots, zones, belts, bright NH$_3$ clouds) whose 
sizes, shapes, and locations vary over time. Although much cooler, Jupiter is our best 
local analogue to ultracool atmospheres and its time--evolution may also serve as a 
first template for interpreting atmospheric dynamics in ultracool atmospheres.

We employed the high temporal cadence of a unique HST/Jupiter spatially resolved ``truth test'' 
imaging data set to validate the recovery/retrieval of ultracool features in spatially unresolved 
exoplanets and brown dwarf atmospheres with our \textit{Aelous} model as described herein. 
With each HST image integrated over the full disk of Jupiter, these imaging data provide a direct 
photometric analog rotational light curve to unresolved point sources (giant exoplanets and brown 
dwarfs) -- but at extremely high photometric SNR ($\sim$30,000 per temporal sample). Importantly, 
these data simultaneously provide unequivocal imaging knowledge of the origin of spatially 
collapsed light curve variations in two spectral bands, thus enabling this validation experiment.
We will make this dataset publicly available via ADS/VIZIR.

Finally, we applied \textit{Aeolus} to two well--studied, rotating brown dwarfs in the L/T transition: 
2M2139 and SIMP0136. We used observations taken by \citet[][]{apai13} using the 
Wide Field Camera 3 on the Hubble Space Telescope. Observations were obtained with the G141 
grism, and \citet[][]{apai13} performed synthetic photometry in the core of the standard 
J-- and H--bands.
We compare our maps with the \textit{Stratos} maps \citep[][]{apai13}, and Fourier maps.

This paper is organized as follows. In Sect.~\ref{sect:mcmc_code} we present 
\textit{Aeolus}. In Sect.~\ref{sect:up_validation} we present our HST data and their 
reduction (Sect.~\ref{sec:hst_d_red} to~\ref{sect:SL}), make a phenomenological analysis of the 
jovian snapshots (Sect.~\ref{sect:phenom_jup}), and analyze the retrieved light curves 
(Sect.~\ref{sect:lc_inspection}). In Sect.~\ref{sect:phenom_jup} we validate \textit{Aeolus} 
on Jupiter light curves and compare our results with Fourier mapping 
results. In Sect.~\ref{sect:bds} we apply \textit{Aeolus} to two well--studied brown dwarfs, 
and compare our results against other mapping techniques. 
Finally in Sect.~\ref{sect:discussion} we present a discussion of our results and our 
conclusions.

\section{\textit{Aeolus}: MCMC mapping of cool atmospheres}\label{sect:mcmc_code}

We present \textit{Aeolus}, a Markov--Chain Monte Carlo code to map the 
top--of--the--atmosphere (per observational wavelength, hereafter TOA) structure of 
brown dwarfs and other ultracool atmospheres. Due to use 
of bayesian inference, an MCMC code can fit input observations with 
high--dimensional models (such as the structure of an atmosphere) 
and can provide more accurate estimates of uncertainties and 
correlations in model parameters than other commonly used methods.

Models of hydrodynamical flows in rotating spheres predict that the largest 
structures in atmospheres are ellipses, 
with major axes parallel to the equator \citep[][]{cho96,cho08}. 
Therefore, following \citet[][]{apai13}, we describe the photospheres of 
our targets, at every pressure level probed, as a sum of a mean atmosphere 
and a set of elliptical spots. We assume that variations in the observed 
flux of a brown dwarf are due to these spot--like features. The number of 
spots is a free parameter. For every spot, \textit{Aeolus} 
fits the position (longitude and latitude), angular size, and 
contrast ratio to the background TOA. Both the limb darkening and the 
inclination of our target atmosphere's equatorial plane to the line of sight 
are currently pre-defined. We assume linear limb darkening. Throughout this paper 
we use a limb--darkening coefficient $c\sim0.5$, as an average value between 
Jupiter's $c_{0.275 \mu m}$ and $c_{0.763 \mu m}$ \citep[][]{teifel76}.

Our model light curves follow 
\citet[][]{kipping12} with elliptical spots that do not overlap. 
We allow the contrast ratio (flux per unit surface of spot to flux per unit surface 
of background TOA) of every spot to vary between 0.01 and 1.5, 
and set the maximum allowed number of spots to 5. 
We finally normalize the model light curve 
in a similar manner to the observational light curves. 

According to Bayes' theorem, the level of confidence in a model \emph{x} 
given observations \emph{d} 
is $p(\emph{x}|\emph{d})$ = $p(\emph{d}|\emph{x})$$p(\emph{x})/p(\emph{d})$ 
\citep[see, e.g.,][]{astroML}, where $p(\emph{d}|\emph{x})$ is the probability we observe 
data \emph{d} given that model \emph{x} is true. 
Since there is no intrinsic reason why \textit{Aeolus} should prefer specific 
values of the parameters it fits (longitude, latitude, size and contrast ratio) over others, 
we make no prior assumptions about the possible values of these parameters, 
and we assign a uniform (i.e., uniformed) prior ($p(\emph{x}) \sim1$) over 
their respective parameter ranges. 
We assume that the observational errors are nearly Gaussian,  
with known variances, 
and adopt a normal likelihood distribution ($p(\emph{d}|\emph{x})\sim$exp[-$\chi^2$(\emph{x})/2]). 

\textit{Aeolus} combines a Gibbs sampler with a Metropolis-Hastings algorithm 
\citep[see, e.g.,][]{chib95,tierney94}, 
using a random--walk Metroplis--within--Gibbs algorithm.
At each step of the MCMC chain we use a Gibbs sampler to vary a random parameter 
(make a ``jump''). A new model light curve is generated using 
the new set of parameters and the latter is accepted or rejected, using a 
Metropolis-Hastings algorithm. The initial--guess light curve's fitness to the observed light curve 
is compared to the fitness of the ``jump'' light curve by comparing the 
probability $P=e^{-(\chi_{jump}^2-\chi_{init}^2)/2}$ to a random number $\alpha$ 
($\alpha\in$[0,1]). If $P\geq\alpha$ the new ``jump'' state is accepted and otherwise 
discarded and a new trial ``jump'' is made using the Gibbs sampler. The process is 
repeated $N$ times, predefined at the start of the chain. To remove biases 
rising from our selection of initial conditions, we remove a 10\% of the chain 
\citep[see, e.g.,][]{ford05}. 

The choice of the best fitting model takes into account the minimization of the 
Bayesian Information Criterion (BIC) \citep[][]{schwarz78}. For a given model \emph{x} 
the Bayesian Information Criterion is BIC$\equiv-2\ln[L^0(\emph{x})]+k\ln N$, where $L^0(\emph{x})$ 
is the maximum value of the data likelihood, $k$ is the number of model parameters, and 
$N$ the number of data points of our observations \citep[for a recent review see, e.g.,][]{astroML}. When two 
models are compared, the one with the smaller BIC is preferred, and if both models have 
the same BIC the model with the fewer free parameters is preferred. 

\begin{figure}
\centering
\includegraphics[height=67mm]{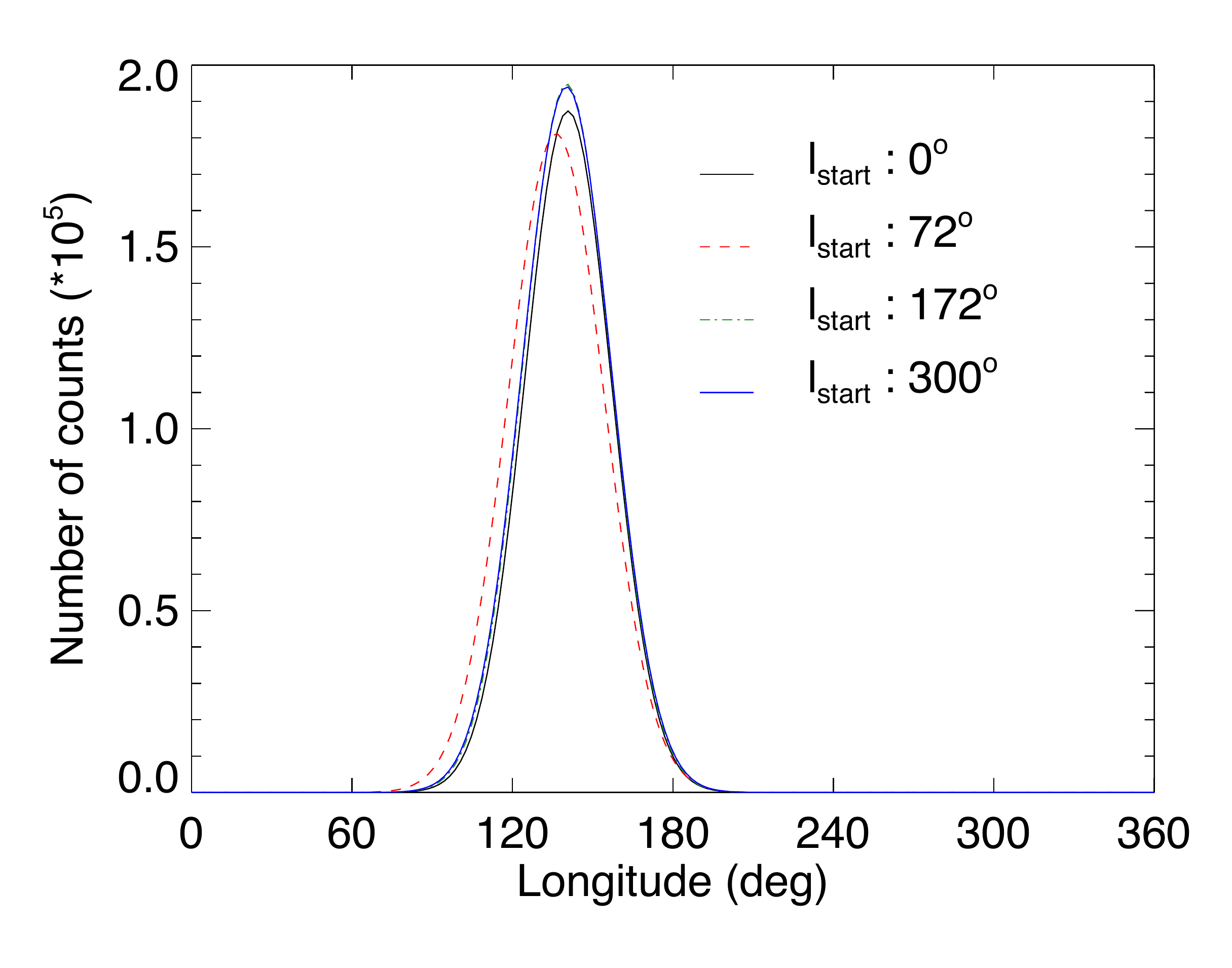} 
\caption{Posterior distributions of four MCMC chains for the longitude of a 
spot of a model atmosphere. The chains start from different locations in the 
longitude--space ($l_{start}=[0^\circ,72^\circ,172^\circ,300^\circ]$), and we control 
their convergence using the $\hat{R}$ criterion. For the four chains shown here 
$\hat{R}\sim$1.01.}
\label{fig:convergence}
\end{figure}

Finally, to control that the solution on which our MCMC chains converge does not 
depend on our initial guesses, we run multiple, independent chains with different 
initial guesses (see, e.g.~, Fig.~\ref{fig:convergence}) and use the Gelman \& Rubin 
$\hat{R}$ criterion to control the convergence of the chains \citep[][]{gelmanrubin92}. 
To accept a solution we check that $\hat{R}$ is always less than 1.2.

We do not include differential rotation or temporal evolution of spots in our code. 
Modeling light curves that vary from one rotational period to the next in \textit{Aeolus} we split
the light curves in rotational periods and fit every partial light curve separately. 
We then compare the successive maps and control whether the retrieved 
variations are physically plausible in the given timeframe.

In the future \textit{Aeolus} will be modified to fit the inclination and limb darkening 
of our targets as free parameters. We will also incorporate temporal evolution of 
features in \textit{Aeolus} in a physically self--consistent manner.

\section{Validating \textit{Aeolus} on Jupiter}\label{sect:up_validation}

A wealth of information exists on Jupiter's atmospheric 
cloud structure and dynamics \citep[see, e.g.,][]{bagenal04,depaterlissauer10}. 
Atmospheric dynamics and a large number of atmospheric features (e.g.,~the 
Great Red Spot (GRS), 5$\mu$m hot spots), indicate that the disk integrated signal of 
Jupiter varies on the rotational timescale (due to rotational modulations, \citealt[see, e.g.,][]{karalidi13} ) 
and on much longer timescales (due to atmospheric circulation). Jupiter's rotational period of  
9$^\mathrm{hrs}$55$^\mathrm{m}$27$^\mathrm{s}$.3 \citep[][]{depaterlissauer10} is 
comparable to that of brown dwarfs \citep[see, e.g.,][]{metchev14}. Clouds in the jovian 
atmosphere, primarily NH$_3$ ice \citep[see, e.g.,][]{west86,simonmiller01}), 
are different from the ones predicted in L to T brown dwarfs (sulfide, Mg--silicate, perovskite 
and corundum clouds)  and the first directly imaged exoplanets \citep[see, e.g.,][]{burrows06,marley02,marley13}.  
They can be comparable though, to the ones in Y dwarfs \citep[][]{morley14b,luhman14} and cooler giant 
exoplanets we directly detect in the future. 
The wealth of variable atmospheric structures, in combination with the ability to get spatially resolved, 
whole--disk images against which we can compare our maps, 
makes Jupiter an ideal target for the validation and 
testing of the sensitivity and limitations of \textit{Aeolus}.

 \begin{figure}
\centering
\includegraphics[height=70mm]{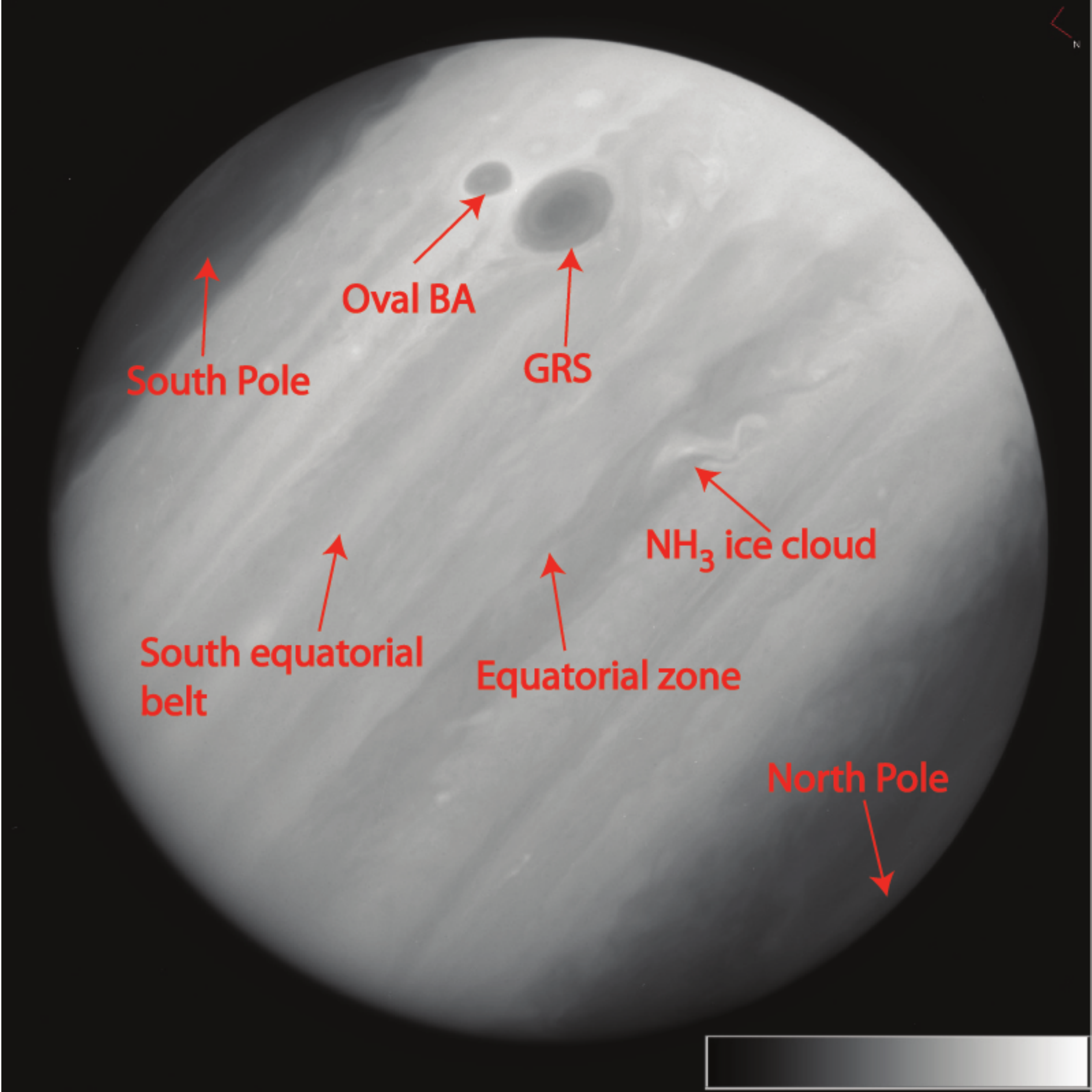}
\centering
\includegraphics[height=70mm]{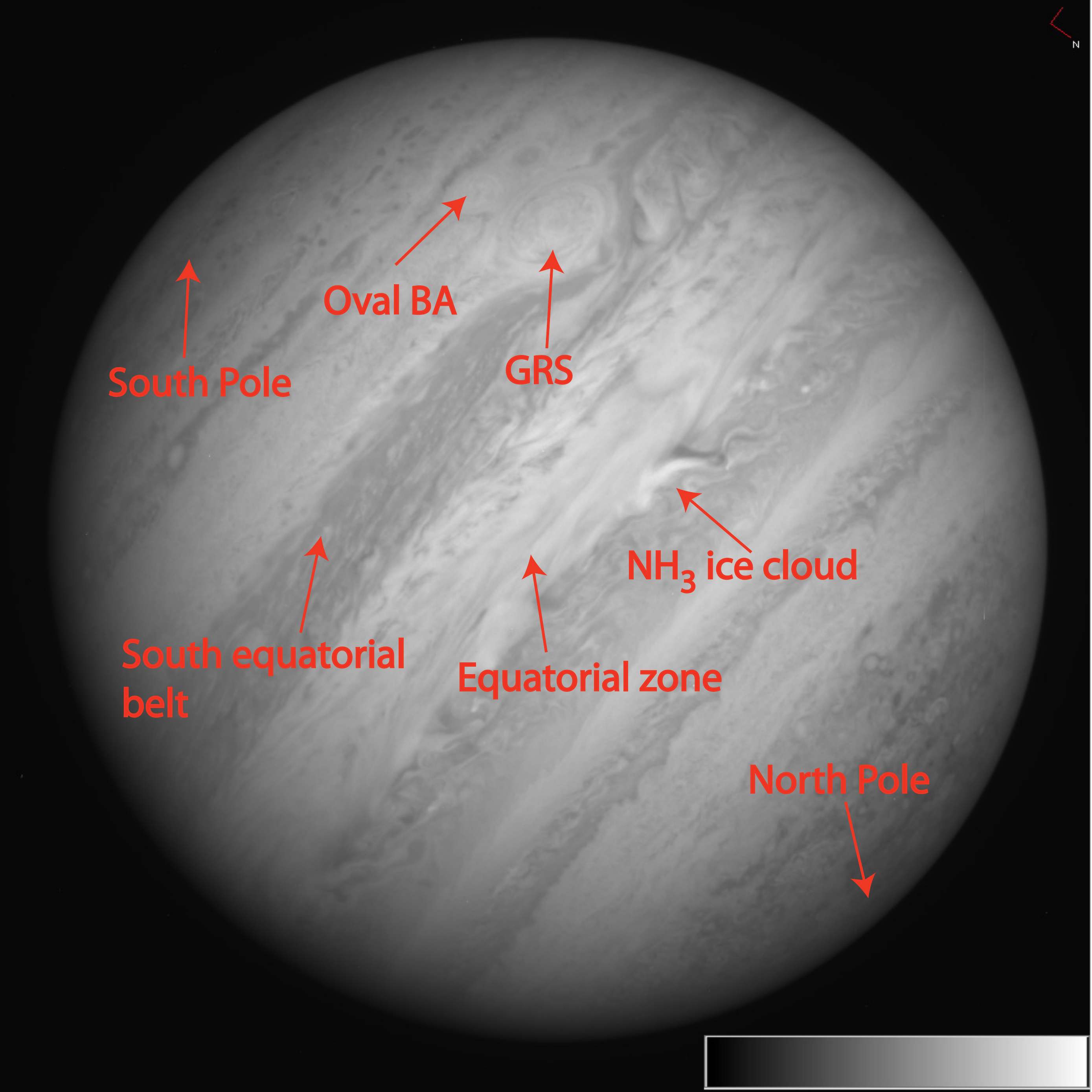}
\caption{Jupiter imaged in the U--band (top panel) and the R--band (bottom panel) at 
a phase of $\sim$0.1. 
The images shown are 1101 x 1101 pixel region extracts from the original 2048 x 2048 
pixel imaging detector sub-arrays centred on the planet. For all images, a 
linear grey-scale  display stretch is used with surface brightness encoded 
as indicated by the scale bars. To optimally tile the full dynamic display range of the data, 
different display scales (in instrumental units of electrons per pixel) are 
used for the F275W and F763M images 
with: F275W--0 (hard black) to 45,000 (hard white) electrons per pixel, 
F763M-- 0 (hard black) to 60,000 (hard white) electrons per pixel. We note 
that different wavelengths probe different layers in the jovian atmosphere and thus the images 
differ considerably (see Sect.~\ref{sect:phenom_jup}). }
\label{fig:jup_phase}
\end{figure}

We applied \textit{Aeolus} to our Hubble Space Telescope's (HST) 
observations of Jupiter. Jupiter was observed with HST Wide--Field Camera 3  (WFC3)
in UTC 19-20 September 2012 during 21.5$^\mathrm{hrs}$, i.e., 2.2 jovian rotations.
Observations were performed in the F275W and 
F763M bands. Data acquisition and reduction is further described in 
Sect.~\ref{sec:hst_d_red}. With their unprecedented high signal--to--noise 
ratio (on average, 26,600 in the F275W and 32,800 in the F763M) full disk photometry of Jupiter, combined with 
high--resolution spatially resolved images over a continuous timespan of more than two jovian days, 
these observations provide us with a unique dataset.  
Various Jovian sub-regions have been studied extensively 
\citep[see, e.g.,][]{simonmiller01,shetty10} and a number of 
full--disk snapshots of Jupiter, Earth and other Solar System planets 
exist \citep[see, e.g.,][]{smith81,cowan09}, but 
to our knowledge there are no previous continuous observations of the full--disk of 
Jupiter or any other Solar System planets. We applied our mapping code on 
these unique light curves and compare the derived maps with the HST images of Jupiter.

 \subsection{HST data \& reduction} \label{sec:hst_d_red}
 
 Time resolved, full disk, photometric UVIS imaging observations of Jupiter, 
 spanning 21.5 hours ($\sim$ 2.2 Jovian rotations), were obtained on UTC 19-20 
 September 2012 with the HST WFC3 (pixel scale $\sim$40 mas pixel$^{-1}$) in 
 HST GO program 13067 (PI: G. Schneider). 
 A total of 124 images were obtained from data acquired in 14 contiguous HST orbits 
 (of $\sim$96 minutes each), sequentially alternating between two spectral filters: F275W 
 (hereafter U--band, $\lambda_\mathrm{pivot}$ = 2704${\AA}$, FWHM = 467${\AA}$) 
 and F763M (hereafter R--band, $\lambda_\mathrm{pivot}$ = 7612${\AA}$ , 
 FWHM = 704${\AA}$). These data were acquired during, and flanking, a transit of Venus 
 as seen from Jupiter \citep[][]{pasachoff12,pasachoff13,pasachoff13b}, with 
 a predictable maximum photometric depth due to 
 geometrical occultation of $\sim$0.01\% (100 ppm), much smaller than the rotation signature 
 from clouds of import to this study.  The potential ``tall poles'' in photometric measurement 
 precision at the levels of possible significance to this investigation are Cosmic Ray (CR) 
 detection and mitigation, instrumental stray light and pointing repeatability.  
 All three are discussed in detail below. We finally corrected our data for the 
 changing Earth--Jupiter and Jupiter--Sun distances, as well as for the changing disk illumination 
 fraction and angular size of Jupiter, over the duration of our observations.

 \subsubsection{Data acquisition} \label{subject:data}
 
 At the time of these observations, the angular diameter of the nearly fully illuminated (99.03\%) 
 disk of Jupiter was $\sim$41.7''. A 2K$\times$2K pixel (80''$\times$ 80'') readout subarray, nominally centered 
 on the planet, was used to reduce readout overheads while also (by its over--sizing) reducing 
 Jovian stray light (encircled energy) escaping the finite imaging aperture field--of--view far from the 
 planet. Exposure times were designed to reach $<$90\% full well depth for the brightest features 
 expected in the Jovian cloud tops (to prevent image saturation, we checked against previous imaging) 
 to yield an aggregate $\sim$2.2$\times$10$^{10}$ electrons combining all $\sim$866,000 WFC pixels in 
 each image that tiled the disk of Jupiter with exposure times T$_\mathrm{exp}$(u) = 29.40 s and 
 T$_\mathrm{exp}$(r) = 0.48 s. Given expected interruptions in data acquisition from Earth occultations, 
 spacecraft south--Atlantic anomaly (SAA) passages (which vary in orbit phase from orbit to orbit), and the 
 instruments' occasional need to pause for an image data ``buffer dump,'' a minimum of six to a maximum of 
 ten images were obtained in each orbit's approximately 54 minute target--visibility period.  When uninterrupted, interleaved 
 intra--image cadences of  225s in U--band, and 214s in R--band, imaging were achieved. Data from the first, 
 and part of the second, HST orbits were (as expected and used for calibration purposes) photometrically 
 partially ``corrupted'' by excess light from Io intruding into the field of view.  Separately, partway through 
 the last (14$^\mathrm{th}$) orbit, the HST pointing control system suffered a guide--star loss--of--lock, degrading the 
 photometric fidelity obtained thereafter.  The photometric data set 
 considered in detail in this paper excludes these degraded data, but are inclusive of all others obtained 
 from UTC 01$^\mathrm{hrs}$21$^\mathrm{m}$43$^\mathrm{s}$ 
 to 20$^\mathrm{hrs}$27$^\mathrm{m}$46$^\mathrm{s}$. The detailed exposure--by--exposure 
 observing plan\footnote{\url{http://www.stsci.edu/hst/phase2-public/13067.pdf}} is available on-line 
 from the Space Telescope Science Institute.

 
\subsubsection{Basic Instrumental Calibration}
The basic (routine) exposure level instrumental calibration of the raw imaging data 
(data set identifier IC3G* in the Mikulski Archive for Space Telescopes 
(MAST)\footnote{\url{http://archive.stsci.edu/hst/search.php}})
 including gain conversion, bias, dark current corrections and flat fielding, was done 
 using STScI's \emph{calwfc3} calibration 
 software\footnote{\url{http://ssb.stsci.edu/doc/stsci$\_$python$\_$2.14/wfc3tools-1.1.doc/html/calwf3.html}} 
 (as implemented in the HST OPUS pipeline). As these raw data were acquired without a need for post--flashing 
 (due to the bright-target field), no post--flash corrections were performed.  Because 
 Jupiter is both a moving, and spatially--resolved rotating target, and data extraction 
 at the full sampling cadence was desired, the individual FLT, not DRZ (``drizzle'' 
 combined) files were used in subsequent post--processing and photometric analysis. 
  
\subsubsection{Astrometric Image Co--Alignment}
Small (few pixel) image offsets were noted in observed images, even those 
using the same guide stars, likely mostly due to imperfections in moving--target tracking. 
Comparable offsets were seen between visits (orbits) where changes in the secondary 
guide stars was required due to the planetary motion. For each filter, all differentially 
imperfectly pointed images were astrometrically co--aligned (registered) prior to the identification 
and subsequent correction of CR affected pixels, and for later large, enclosing aperture, 
photometry. Differential image decentrations were determined from sequential image pairs by 
minimizing the variance in a small (few pixel) width annulus enclosing the limb of Jupiter in difference 
images with iterative ``shifting'' of the image treating ($\Delta$x, $\Delta$y) as free parameters. 
``Shifting'' (with each iteration re--referenced to the original image) was done by sub--pixel 
image remapping via bi--cubic interpolation apodized by a sinc function of kernel width appropriate 
to each filter to suppress ringing. The then astrometrically co--registered FLT files were not 
additionally corrected for the WFC3 geometrical distortion, which is actually preferable to omit 
for high--precision differential photometry in obviating additional flux-density interpolation errors 
in geometrical correction associated pixel remapping.  (N.B.: This is why, by chance of observational 
geometry/spacecraft orientation, geometrically uncorrected FLT images of Jupiter look quite 
round, rather than oblate, as exampled in Fig.~\ref{fig:jup_phase}). 
 
\subsubsection{Cosmic Ray rejection}\label{subject:data_cr}
 Although exposure times (and so susceptibility to CR hits) are small, (multiple) high--energy 
 CR events could photometrically bias even large-aperture photometry if not at least 
 partially mitigated by CR detection and compensation  Since Jovian image structure is 
 not static, the simple oft--used two--image minimum, or multiple, image median approach 
 for intrinsically invariant images is not appropriate.  Here we adopted a hybrid approach, 
 different in process for the sky background region (which includes instrumentally 
 scattered planetary light, so is necessary to correct and later photometer) and for the 
 planetary disk. On disk we use local median spatial filtering, and off--disk we 
 use simple image--pair anti--coincidence detection. 
 
The on--disk region has spatially and temporally variable cloud structure that, even on 
small spatial scales, has detectable changes from image to image at WFC3 resolution even 
at the shortest sampling timescales. Most of these are correlated in two dimensions over at 
least several pixels, whereas CR hits are usually isolated to single pixels or are ``trails'' only one 
pixel in width.  Thus, we identify most CR--affected pixels as outliers identified from high--pass 
spatially--filtered images.  Spatial filtering is simply done, for each image, by subtracting a 3$\times$3 
pixel boxcar image convolution of the image from image itself.  On--disk CR--corrupted pixels are 
then identified from the spatially--filtered images as $\geq$+3.5$\sigma$ outliers  w.r.t. 1$\sigma$ 
deviations in an on--disk 700$\times$700 pixel planet--centered sub--array fully circumscribed 
by the disk of Jupiter. (In detail, with experimentation using different size filtering kernels we found in 
the 3$\times$3 case $<$3$\sigma$ erroneously finds pixels that are correlated with disk structure, 
and $>$ 4$\sigma$ ``misses'' many uncorrelated pixels (tested by injecting CR--like signals into 
template images). While the surface brightness of the disk is locally variable, a constant 3.5$\sigma$ threshold 
w.r.t. the (centrally brighter) 700$\times$700 pixel disk--centered subarray provides a statistically 
uniform clipping (identification) level w.r.t. CR energy (intensity) for all images (in each filter). For the full 
ensemble of U and R images, respectively, the medians of the 1$\sigma$ deviations in the central 
700$\times$700 pixel regions are uniformly adopted to establish the clipping threshold for 
all like--filter images:  U$_\mathrm{median}$(1$\sigma$) = 212.6 counts/pixel,  
R$_\mathrm{median}$(1$\sigma$) = 255.6 counts/pixel (compare full--disk averaged signal 
levels $\sim$30,000 counts/pixel and $\sim$2.2E10 counts integrated over the full disk of Jupiter.

Off--disk (sky) CR--compromised pixels are found (to a limiting threshold) in a two--step 
process. Step 1: In each visit, sequential image pairs in the same filter are inter--compared to find 
the smaller--valued of two co--located pixels (for all sky pixels) with the presumption of intrinsic 
background sky image stability between same--filter sequential images.  In the absence of 
CR events (and sky instability) the sky--region images will differ significantly only by 
instrumental noise plus photon noise in the background.  The smaller--valued of  each of the 
two--pixel pairs is used to assess the sky background at that pixel location. 
Step 2: In infrequent cases where independent CR--events may pollute the same 
pixel in sequential images this method will fail to find a proper sky estimation for that pixel. 
Those pixels are then identified by  sigma--clipping against the local background after pixel--pair minimization.
The spatially mutually--exclusive on and off disk regimes are then re--combined to produce a ``CR cleaned'' 
image to the above detection threshold limits.

\begin{figure}
\centering
\includegraphics[height=45mm]{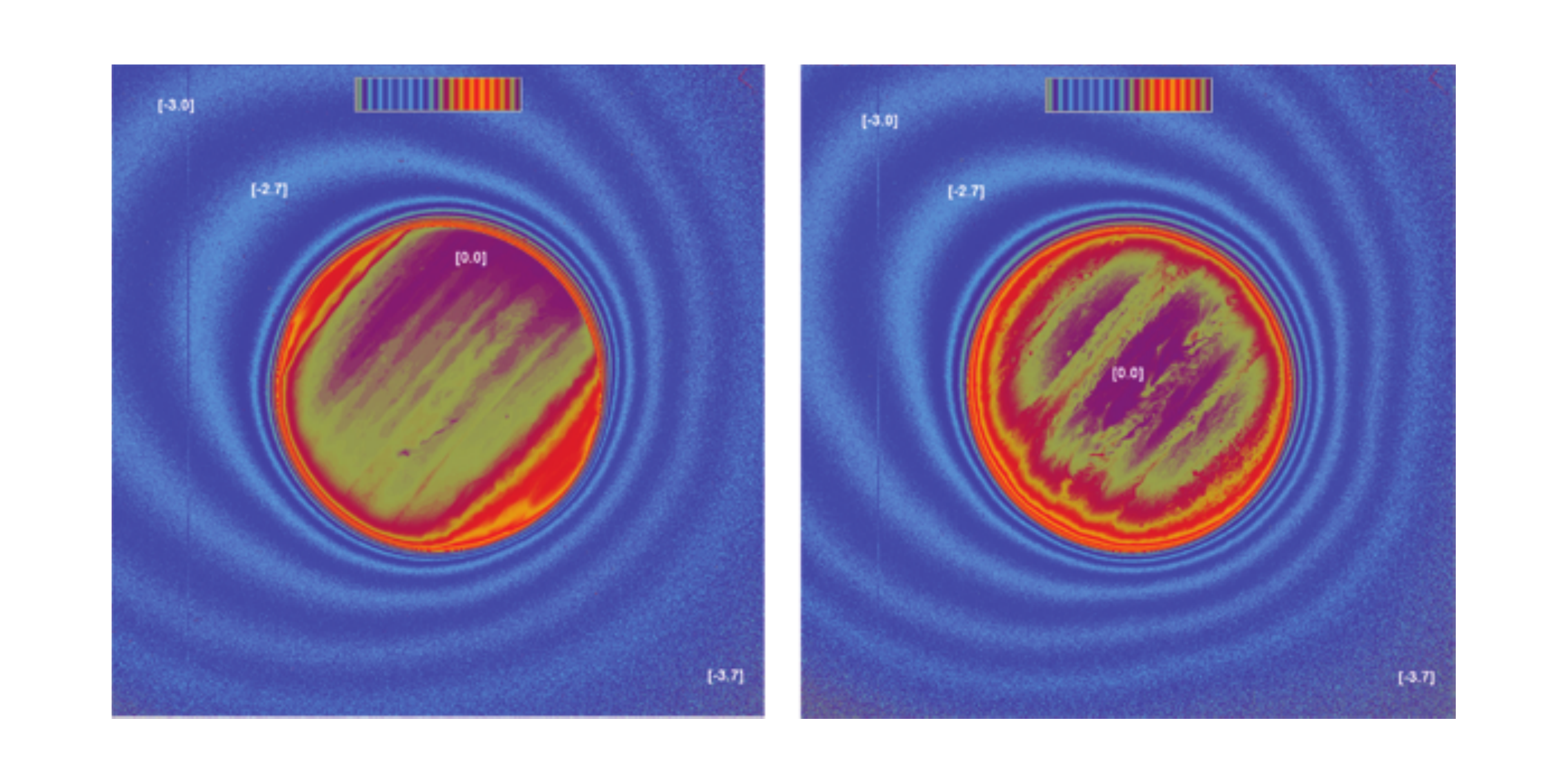}
\caption{Representative circumplanetary stray/scattered light.  
Left: U-band, Right: R-band.  Log 10 display normalized to peak on--disk intensity. 
$+3.3$\% isophotes in log10 space from [-4] to [0] dex counts/pixel.}
\label{fig:scatlight}
\end{figure}

 \begin{figure}
\centering
\includegraphics[height=70mm]{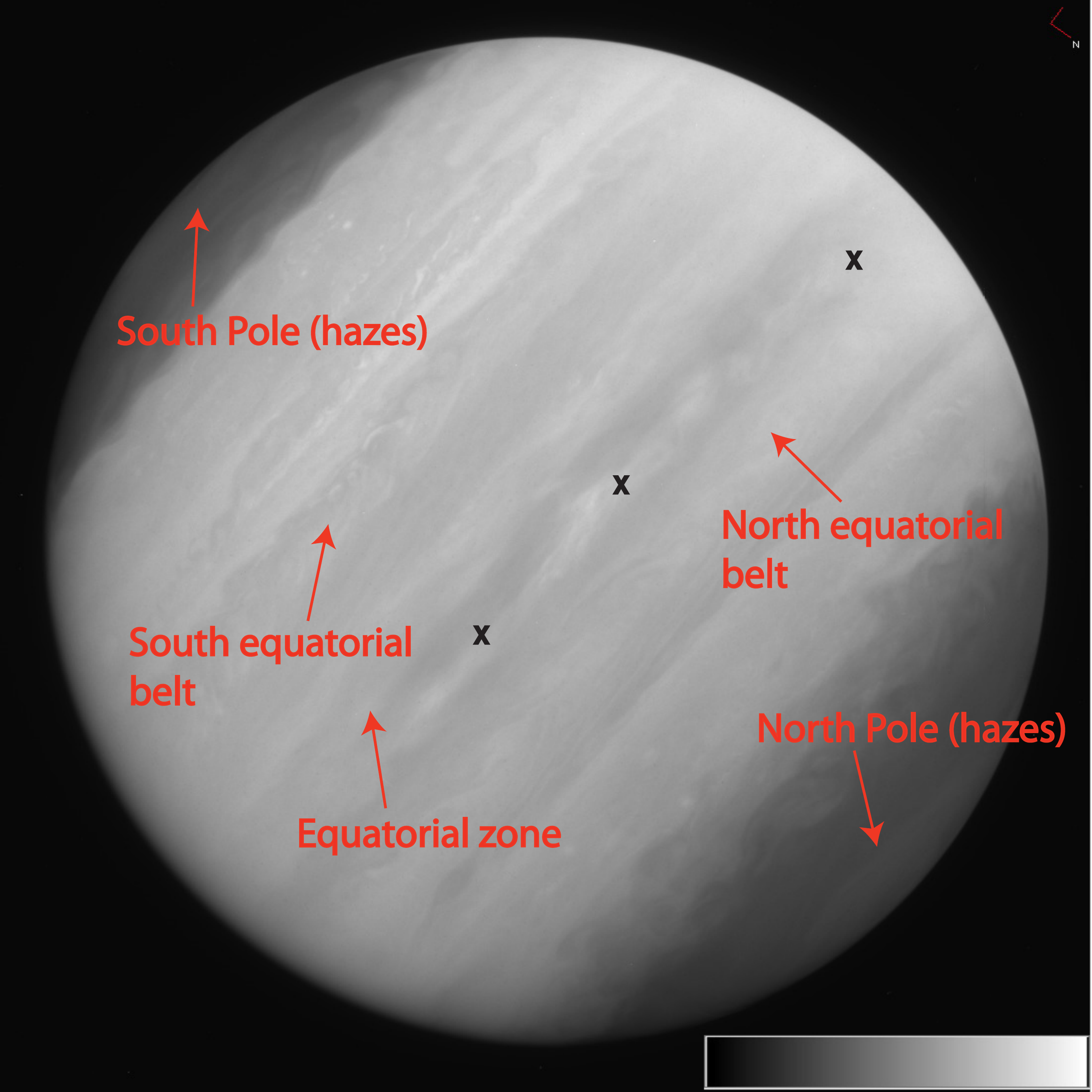}
\centering
\includegraphics[height=70mm]{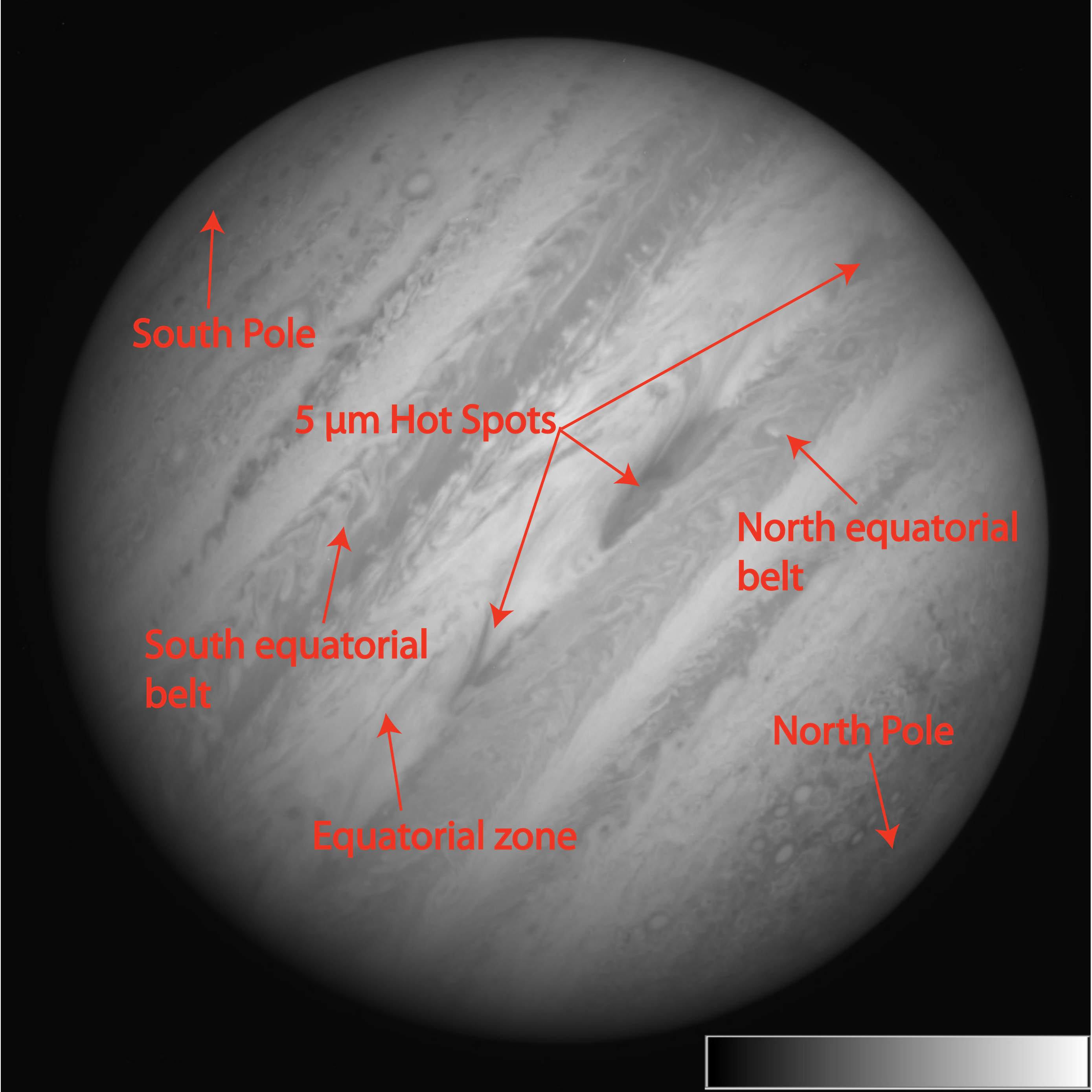}
\caption{U--band (top panel) and R--band (bottom panel) Jupiter snapshot at a 
phase angle of $0.3$. We note that the 5 $\mu$m hot spots we see in the R--band have no 
visible counterpart in the U--band (locations marked with X). 
Colorbars as in Fig.~\ref{fig:jup_phase}.}
\label{fig:jup_images}
\end{figure}

\subsubsection{Instrumental (Stray) Light and HST Pointing Authority in Detail} \label{sect:SL}
 Instrumentally scattered light from the large, bright, disk of Jupiter into the circumplanetary 
 sky background is both circularly asymmetric and falls off much more slowly (as expected) than the 
 PSF halo of an isolated point source.  This is shown illustrated in Fig~\ref{fig:scatlight} for representative 
 F275W and F763M brightness maps shown as contour images log10 stretched normalized to the 
 surface brightness of the brightest parts of the Jovian disk.  As can be seen, at the edge of the FOV 
 the ``sky'' brightness from Jovian stray light has declined to only $\sim10^{-3}$ to $10^{-4}$ of the 
 peak surface brightness of the disk. (The full dynamic display range in this image display is 
 [-4] to [0] dex relative to the brightest parts of the disk). 
 
The ability to achieve high--precision source--enclosing large--aperture (including sky) differential 
photometry depends, then, upon the stability of the stray light pattern, i.e., if the planet moves in the 
2k$\times$2k imaging subarray between exposures, the stray light pattern will shift. Its structure may 
then change resulting in different amounts of stray light falling out of the photometric aperture used, not only 
because of decentration (that is post--facto compensated; see Astrometric Image Co--Alignment), but from 
a possible change in the two--dimensional structure of the scattered light pattern with target displacement 
in the FOV.  HST pointing stability while using two Fine Guidance Sensor fine lock guiding  (used for these 
observations) with respect to the planetary tracking precision is approximately 4 mas RMS. 
Target re--acquisition precision, with 
the same guide stars in successive orbits, is $\sim$10 mas or better from visit to visit.  Fortuitously, the same 
primary guide star (which is used for attitude control) was available and used for all 14 visits. 
Because of Jupiter's motion through the sky, however, the observing program switched twice to different 
secondary guide stars (which are used for roll control).  Re--using the same primary guide star for all visits 
should (to close to first order) result in the target (center of Jupiter) placement in the aperture very repeatable 
in all visits, but a small differential roll error (tenths of a degree) between Visits 07 and 08, and again Visits 
12 and 13, when the secondary guide stars were switched, could potentially bias the aperture photometry 
(with an undersized aperture) -- but is not seen in these data when reduced (masking aperture edges) and measured.

\subsection{Phenomenological analysis of Jupiter images}\label{sect:phenom_jup}

Identifying the most prominent features in Jupiter's U-- and R--band 
images is important for interpreting the jovian light curves and 
controlling the validity of our mapping technique. In Fig.~\ref{fig:jup_phase} we
present a Jupiter snapshot in the U (top panel) and R (bottom panel) bands 
at a rotational phase angle of $\sim$0.1. 
Note that the images are oriented with the South pole located 
on the upper left corner of the images.

Even though the two images are taken at the same rotational phase angle 
they differ considerably. In the U--band the 
jovian disk is nearly homogeneous (jovian zones and bands appear smooth 
and of comparable intensity) and the most prominent features are the Great Red Spot 
(GRS) and Oval BA (see Fig.~\ref{fig:jup_phase}). 
Additionally, the jovian poles appear darker than 
the central parts of the disk. On the other hand, in the R--band 
the GRS and Oval BA disappear, i.e., they have the same color and intensity as the 
South temperate belt. The jovian disk appears clearly heterogeneous due 
to the prominent zones and belts, while the poles appear darker due only to 
limb darkening. This is due to the different atmospheric layers 
probed at the two wavelengths. 

In particular, the short--wavelength 
U--band probes the higher jovian atmosphere down to $\sim$400 mbar \citep[][]{vincent00}, 
and we can observe the GRS (top pressure of $\sim$250 mbar) and the 
Oval BA (top pressure $\sim$220 mbar) \citep[][]{simonmiller01}. 
The zones and belts, on the other hand, 
have cloud-top pressures of 600 mbar down to 1 bar \citep[][]{simonmiller01}, 
making them visible at the longer wavelength (R--band) observations, which 
probes deeper pressure levels in the atmosphere down to $\sim$2 bars \citep[][]{irwin03}. 

Stratospheric hazes cover Jupiter's poles (at pressures of 
10--100 mbar), consisting of aggregates of particles that are small in 
comparison to the incident light \citep[][]{west91,ingersoll04}. These hazes are 
thought to be condensed polycyclic aromatic hydrocarbons or 
hydrazine, generated in the upper stratosphere from CH$_4$ under the 
influence of the solar ultraviolet radiation \citep[][]{friedson02,atreya05}. Due to their 
high altitude we expect the polar hazes to be visible in the U--band observations 
and not in the R--band. Additionally, we expect them to appear darker than 
the background NH$_3$ clouds \citep[e.g.,][their Fig.1]{karalidi13}, as we indeed 
see in Fig.~\ref{fig:jup_phase}.

Jupiter's GRS is located at a latitude of $22.4^\circ\pm0.5^\circ$S, with a 
latitudinal extent of $11^\circ\pm1^\circ$ \citep[][]{simonmiller02}  and a longitudinal 
extent of $18.07^\circ\pm0.91^\circ$ as of 2000 \citep[][]{trigo00,simonmiller02}, which given 
a linear shrinkage rate of $-0.114^\circ/\mathrm{yr}$  \citep[][]{simonmiller02},
would translate to $16.70^\circ\pm0.91^\circ$ in 2012.
Oval BA is located at $33^\circ$S latitude \citep[][]{wong11}, and 
extends $\sim5^\circ$ in latitude and $\sim11^\circ$ in longitude \citep[][Fig.18]{shetty10}.

In Fig.~\ref{fig:jup_images} we present Jupiter at a phase angle of 0.3. In the R--band 
(right panel) we notice the existence of a large hot spot on the North hemisphere. 
In hot spots, the atmospheric cloud content is low and the heat can escape from deeper 
layers without much absorption. Hot spots thus appear dark in the visible, but bright 
at 5$\mu$m \citep[][]{vasavada05}. Jupiter's hot spots are centered 
around $6.5^\circ$N--$7^\circ$N latitude \citep[][]{ortiz98,simonmiller01}. 
Their longitudinal to latitudinal extent ratio varies between 1:1 to 7:1, while 
strong zonal flows at the north and south boundaries of these features 
limit their latitudinal size to a maximum of $8^\circ$ \citep[][]{choi13}. The hot 
spot of Fig.~\ref{fig:jup_images} has a latitudinal extent of $\sim4^\circ$ and 
a longitudinal extent of $\sim18^\circ$.
 
 \subsection{Light curve inspection} \label{sect:lc_inspection}
 
 In Fig.~\ref{fig:jup_lightcurves} we present the normalized R 
 (red boxes) and U (blue circles) band HST light curves of Jupiter. 
 Before testing our mapping code we inspected the light 
 curves and compared them with the HST images of Jupiter.

 The R--band light curve has a peak--to--peak amplitude of 
 $\sim$2.5\% and appears to be a smooth sinusoidal function. 
 In comparison, the U--band light curve has a small peak--to--peak 
 amplitude of $\sim$0.5\% and its small scale structure indicates that it is influenced by 
 multiple atmospheric structures. 

A comparison of the R--band light curve with HST images shows that the hot spot  
of Fig.~\ref{fig:jup_images} (left panel) is responsible for the troughs of 
the light curve (see also Fig.~\ref{fig:jup_lightcurves}), while the GRS for the 
peaks. In the U--band the GRS and Oval BA (see right panel of 
Fig.~\ref{fig:jup_images}) appears to be responsible for the lower flux 
around a phase of 0.9 and 1.8, while the overall small--scale structure seems 
to be due to changes in the distribution of high NH$_3$ ice clouds (see 
also Fig.~\ref{fig:jup_lightcurves}).

\begin{figure}
\centering
\includegraphics[height=67mm]{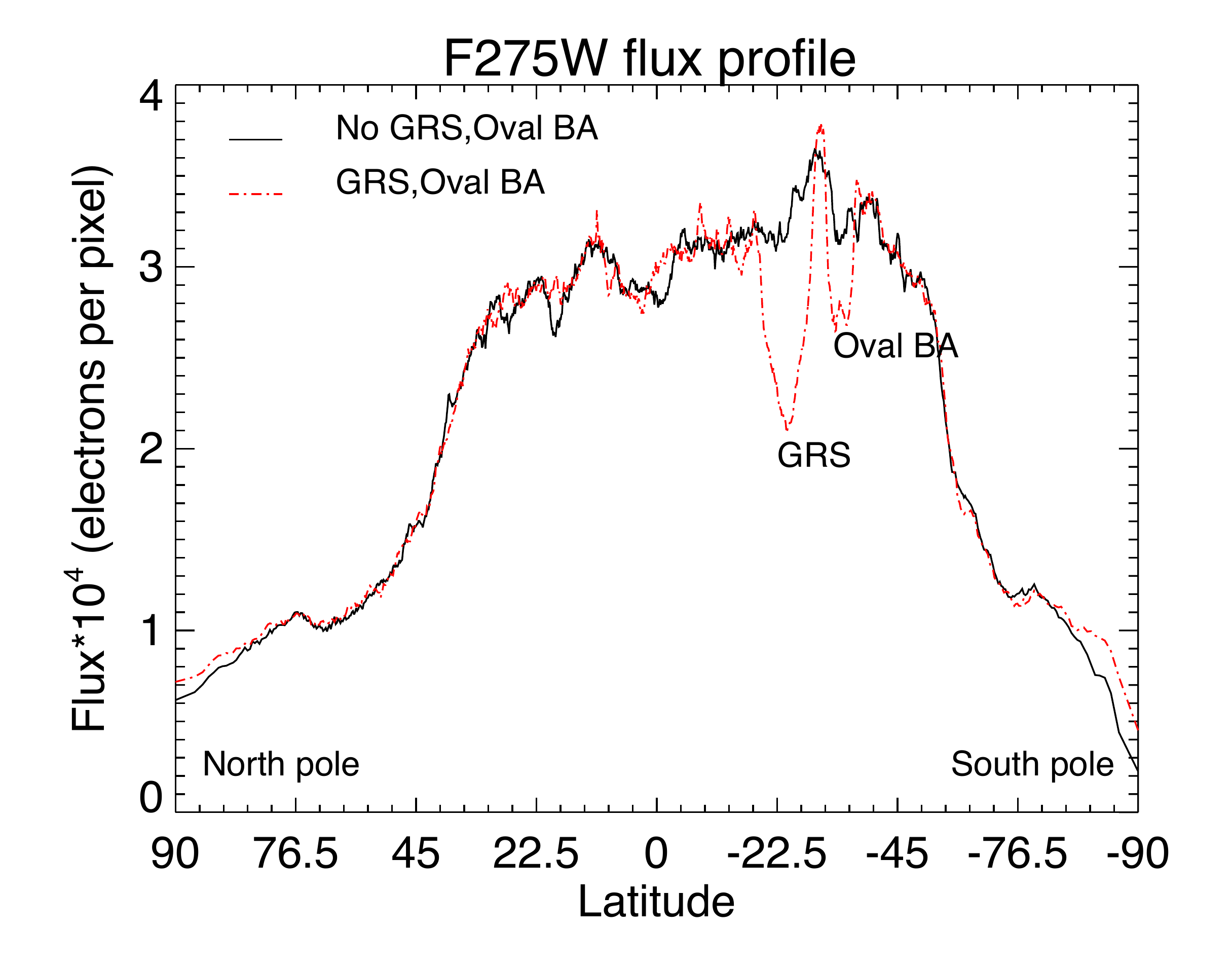} 
\centering
\includegraphics[height=67mm]{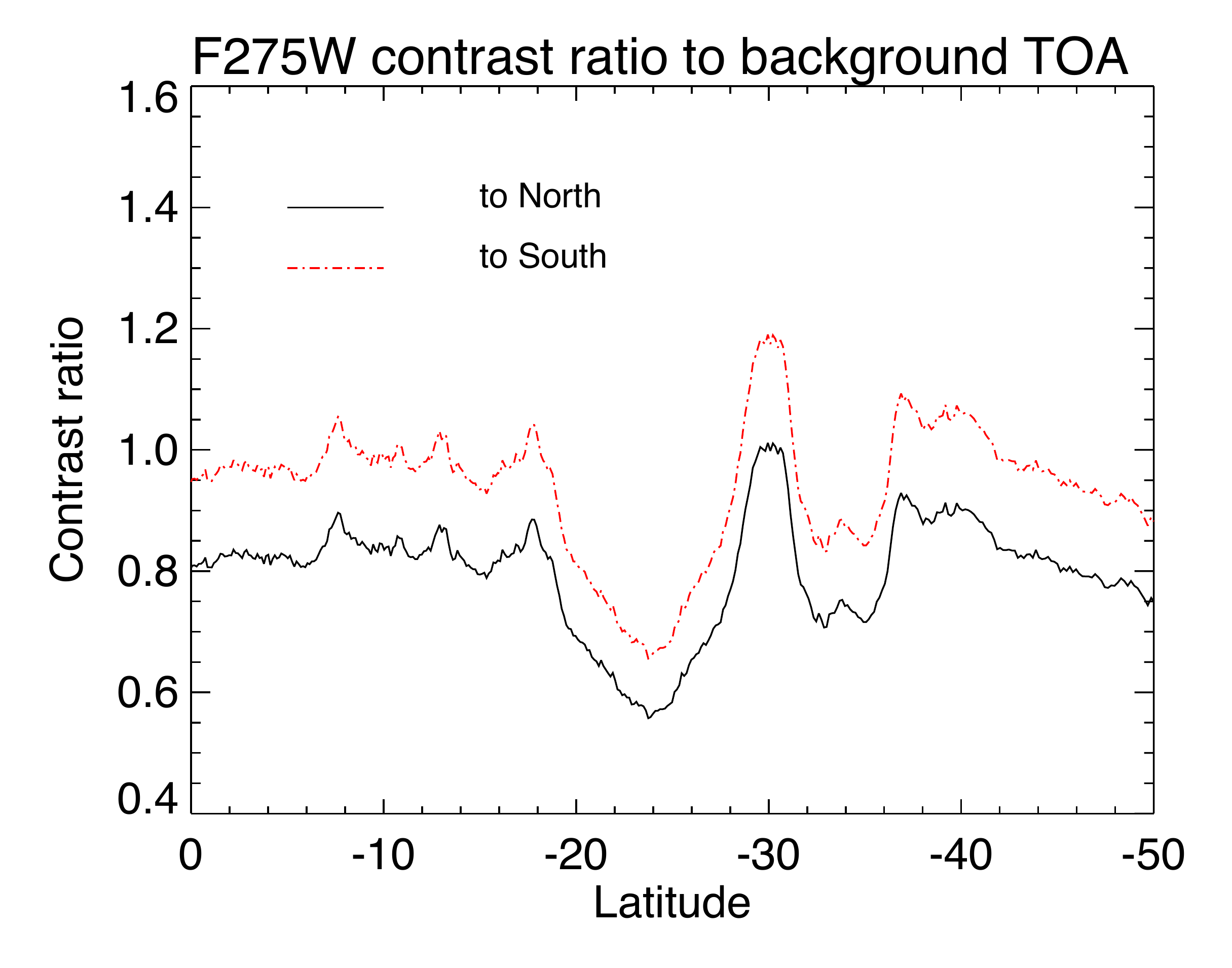} 
\caption{Top: Jupiter latitudinal U--band flux profiles at a longitude of $\sim116^\circ$, 
not passing through the GRS and Oval BA (black, solid line) and 
of $\sim320^\circ$, passing through the GRS and Oval BA (red, dashed--dotted line). 
We define as $0^\circ$ longitude the center of the first 
image acquired during these HST observations. 
We notice that the GRS and Oval BA are darker than their directly 
surrounding disk. Bottom: Contrast ratio of a slice of the top figure to a 
location to the North ($\sim$-17$^\circ$; black, solid line) and to 
the South ($\sim$-29$^\circ$; red, dashed--dotted line) of the GRS.}
\label{fig:jup_profiles}
\end{figure}

\begin{figure*}
\centering
\includegraphics[width=.85\textwidth]{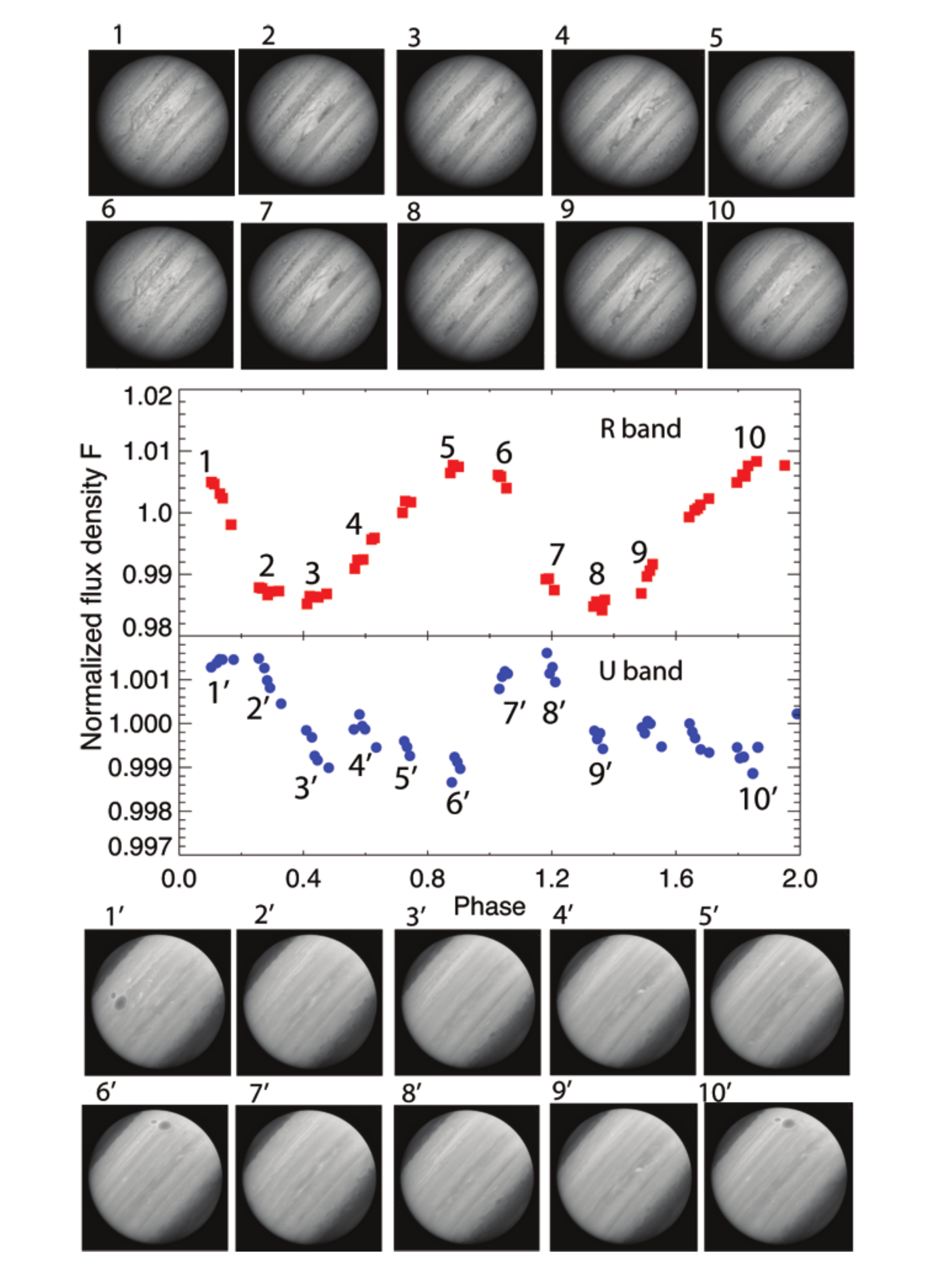}
\caption{Normalized R (red) and U (blue) band light curves of Jupiter. 
The uncertainties in the relative, disk--integrated, photometric measures (each point) 
are estimated as 1$\sigma\leq$ 0.022\%$\pm$0.009\% of the 
measured signal in either filter band. Corresponding snapshot images 
of Jupiter in the R--band (top) and U--band (bottom) are shown for helping 
the reader interpret the light curves. }
\label{fig:jup_lightcurves}
\end{figure*}

We define as $0^\circ$ longitude the center of the first 
image acquired during these HST observations. In Fig.~\ref{fig:jup_profiles} 
(top panel) we show a latitudinal flux profile of Jupiter at a longitude 
of $\sim334^\circ$, passing through the GRS and Oval BA (red, dashed--dotted line) and at a longitude 
of $\sim116^\circ$ (black, solid line). The GRS and Oval BA (around a latitude of $-23^\circ$ and $-34^\circ$ respectively) 
are darker than their surrounding TOA. In particular, the GRS at its darker 
part has a contrast ratio of 0.55 (0.62) to the disk at its north (south) side and Oval BA 
has a contrast ratio of 0.70 (0.79) to the disk at its north (south) side (see bottom panel of Fig.~\ref{fig:jup_profiles}).
Full disk photometry of our images though shows that Jupiter's GRS has a contrast ratio 
of 0.97 to the integrated background jovian disk (as seen in the U--band) and the Oval BA 
has a contrast ratio of 1.17. This is due to the extremely dark poles 
of Jupiter in the U--band. Finally, the big hot spot we see in 
the left panel of Fig.~\ref{fig:jup_images}  has a contrast ratio (as seen in the R--band) of 1.15.

\subsection{Application of \textit{Aeolus}} \label{sect:map_jup}
 
We initially applied \textit{Aeolus} to Jupiter's R--band light curve. We ran $8$ 
chains of length 5,000,000 each, with different initial conditions. We used the 
Gelman and Rubin (\^{R}) criterion to test our chains' convergence. 
Since the light curve shows evolution from one rotation to the next, 
we split it and ran our MCMC code on each rotation 
($10^\mathrm{hr}$ intervals) separately. 

\begin{figure}
\centering
\includegraphics[height=60mm]{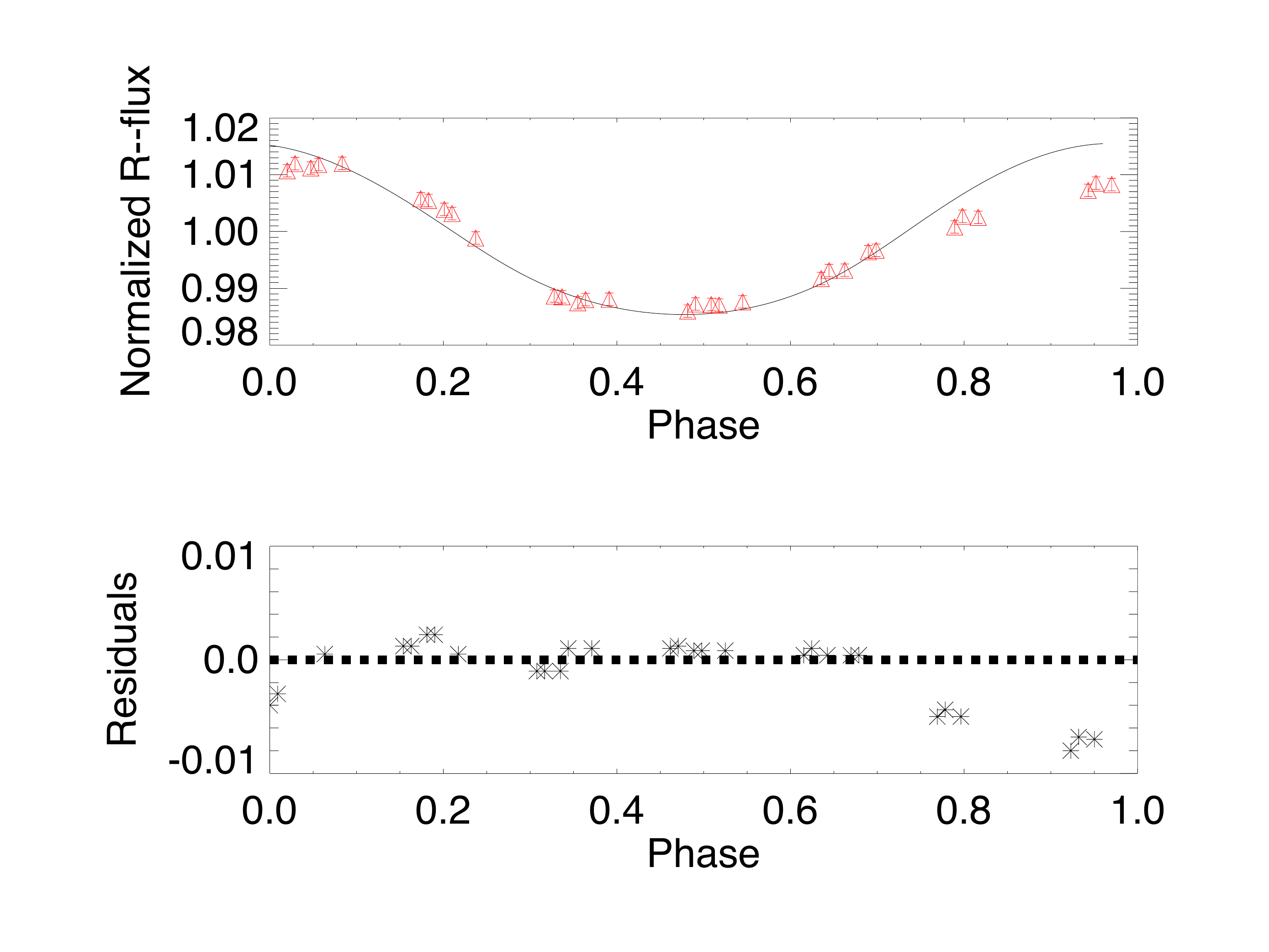}
\caption{Normalized R--band light curve of Jupiter (red triangles) with 
error bars, and best fit \textit{Aeolus} curves (black solid line) for the first rotation (top panel); and  
corresponding residuals (bottom panel).}
\label{fig:jup_fits}
\end{figure}

For the first rotation, we retrieved 2 spots (BIC 19.3) 
located at a longitude of $128.8^\circ\pm12.8^\circ$ and $312^\circ\pm10^\circ$; 
a latitude of $23^\circ\pm12^\circ$ and $31^\circ\pm16^\circ$;
with a size of $16.7^\circ\pm1.8^\circ$ and $18^\circ\pm4^\circ$; 
and a contrast ratio of $0.96\pm0.20$ and $1.2\pm0.2$.
For the second rotation, we retrieved 2 spots (BIC 15.6), 
located at a longitude of $126^\circ\pm15^\circ$ and $315^\circ\pm14^\circ$; 
a latitude of $22^\circ\pm12^\circ$ and $31^\circ\pm16^\circ$;
with a size of $18^\circ\pm3^\circ$ and $18^\circ\pm4^\circ$; 
and a contrast ratio of $1.04\pm0.18$ and $1.2\pm0.2$.
The \textit{Aeolus}--retrieved spot properties are, 
within the error bars, in agreement with the properties of the hot spot and 
the GRS as presented in Sect.~\ref{sect:phenom_jup}. 
For completeness, in Fig.~\ref{fig:jup_fits} we show the normalized 
R--band light curve (red triangles) with error bars, and the best fit 
\textit{Aeolus} model (black, solid line) for the first rotation (top panel), and 
the residuals (bottom panel).

We then applied \textit{Aeolus} to Jupiter's U--band light curve. The 
U--band light curve has a smaller amplitude and its temporal evolution 
is more pronounced than that of the R--band. We again 
split the curve into two rotations and fit each curve separately. 
 
For the first rotation, \textit{Aeolus} retrieved 1 
spot (BIC 24.5 vs 28.7 for two spot model) located 
at a longitude of $290^\circ\pm20^\circ$ and 
a latitude of $24^\circ\pm8^\circ$, with a size of $19.6^\circ\pm2.1^\circ$ and a contrast 
ratio of $1.05\pm0.08$ to the background. For the second rotation, \textit{Aeolus} 
retrieved 1 spot (BIC 19.1) 
located at a longitude of $319^\circ\pm14^\circ$ and 
a latitude of $13^\circ\pm7^\circ$, with a size of $20.0^\circ\pm1.0^\circ$ and a contrast 
ratio of $1.22\pm0.14$ to the background. Within the error bars, our retrieved 
spot properties agree with the GRS properties as presented in Sect~\ref{sect:phenom_jup}. 
Note that the latitudinal location and size of the retrieved GRS are slightly offset, due 
to the influence of the Oval BA.

The error in the estimated latitude is large (relative to the mean). 
This is due to the latitudinal degeneracy maps based on 
flux observations present \citep[see, e.g.,][]{apai13}. As expected, 
rotationally homogeneous features such 
as the belts and zones of Jupiter do not leave a 
clear trace in the light curves \citep[see, e.g.,][]{karalidi13}. 
Finally, we note that the Oval BA accompanying the GRS cannot be 
retrieved as a separate feature by \textit{Aeolus}, which is again 
due to the latitudinal degeneracies.

We should note here, that \textit{Aeolus} had difficulties converging, 
given the very small uncertainties of our Jovian light curves. 
\textit{Aeolus} was designed to reproduce simple surface brightness maps 
of ultracool atmospheres, assuming that all heterogeneities on the TOA are elliptical. 
A closer look at the U--band light curves of Fig.~\ref{fig:jup_fits} though, indicates 
that due to the high SNR ratio of our dataset, the light curve shape is also influenced 
by non--elliptical, fine structures, such as high NH$_3$ ice clouds. Since 
the modeling of such fine structure is beyond our scope and \textit{Aeolus}' design, 
and for achieving fast convergence, we increased $\sigma$ by a 
factor of $\sim$4. Doing so, we kept $\sigma$ well below the uncertainties of the highest--precision 
brown dwarf observations \citep[see, e.g.,][]{apai13,yang15}, and allowed \textit{Aeolus} to 
map the major non-rotationally--symmetric features of Jupiter in the U-- and R--bands.

\subsubsection{Fourier mapping of Jupiter} \label{sect:four_jup}

We then compared \textit{Aeolus} maps with those produced using 
Fourier mapping, a commonly used mapping technique in the literature.
Following \citet[][]{cowan08}, and given that our problem was under--constrained, 
we defined the longitudinal brightness map of any planet as: 
$M(\alpha)=\frac{a_0}{2}+\frac{2b_1}{\pi}\cos(\alpha)-\frac{2c_1}{\pi}\sin(\alpha)+
\frac{3b_2}{2}\cos(2\alpha)-\frac{3c_2}{2}\sin(2\alpha)$, 
where $\alpha$ is the angle of rotation of Jupiter or a brown dwarf around its axis. 

\begin{figure}
\centering
\includegraphics[height=60mm]{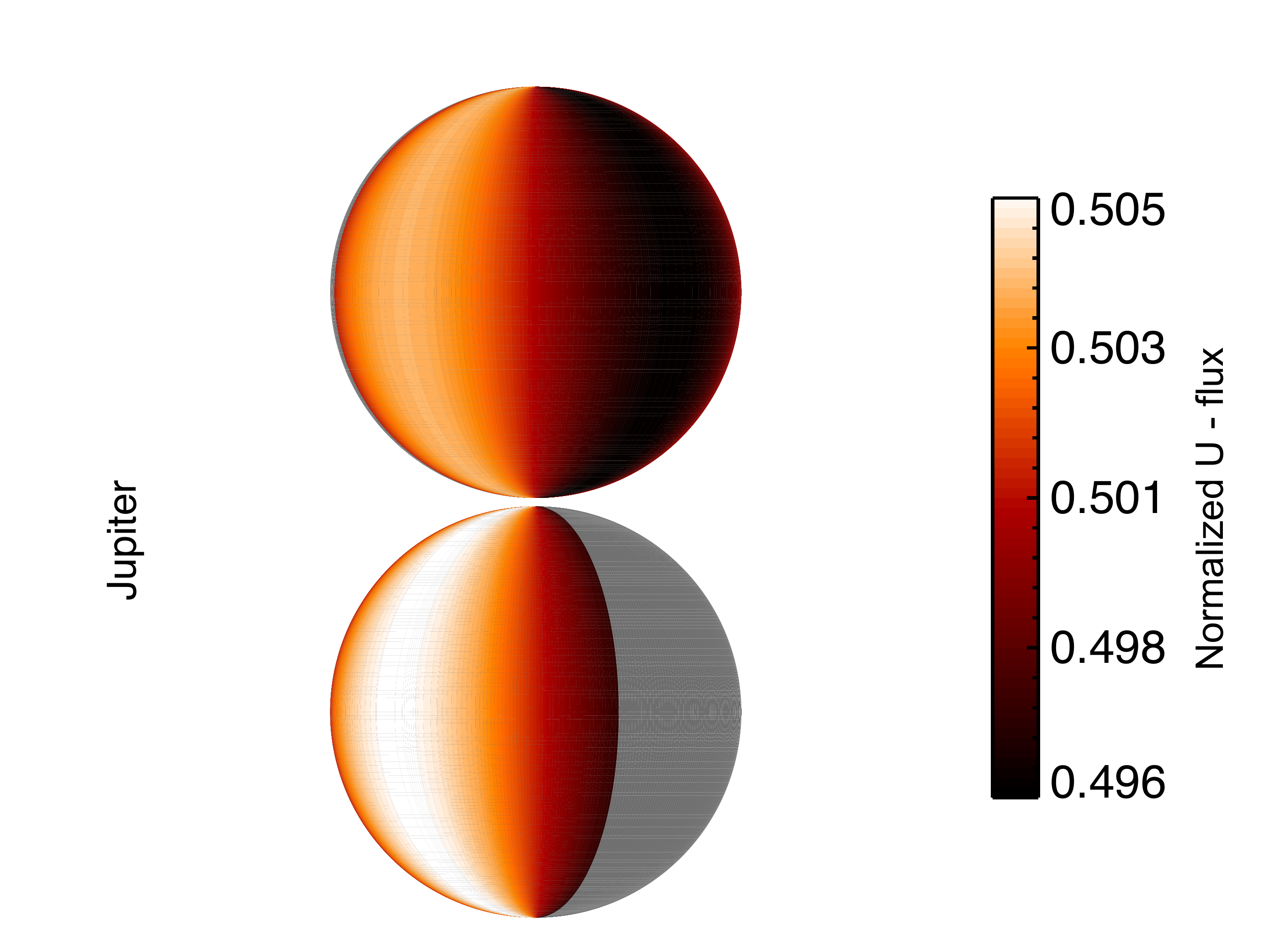}
\caption{Fourier surface brightness map of Jupiter based on its U--band, first rotation.
Upper panel: map centered at a longitude of $140^\circ$. Lower panel: map centered 
at a longitude of $320^\circ$. Grey areas correspond to missing data due to 
an Europa intrusion.}
\label{fig:fft_map_u_jup}
\end{figure}

\begin{figure}
\centering
\includegraphics[height=60mm]{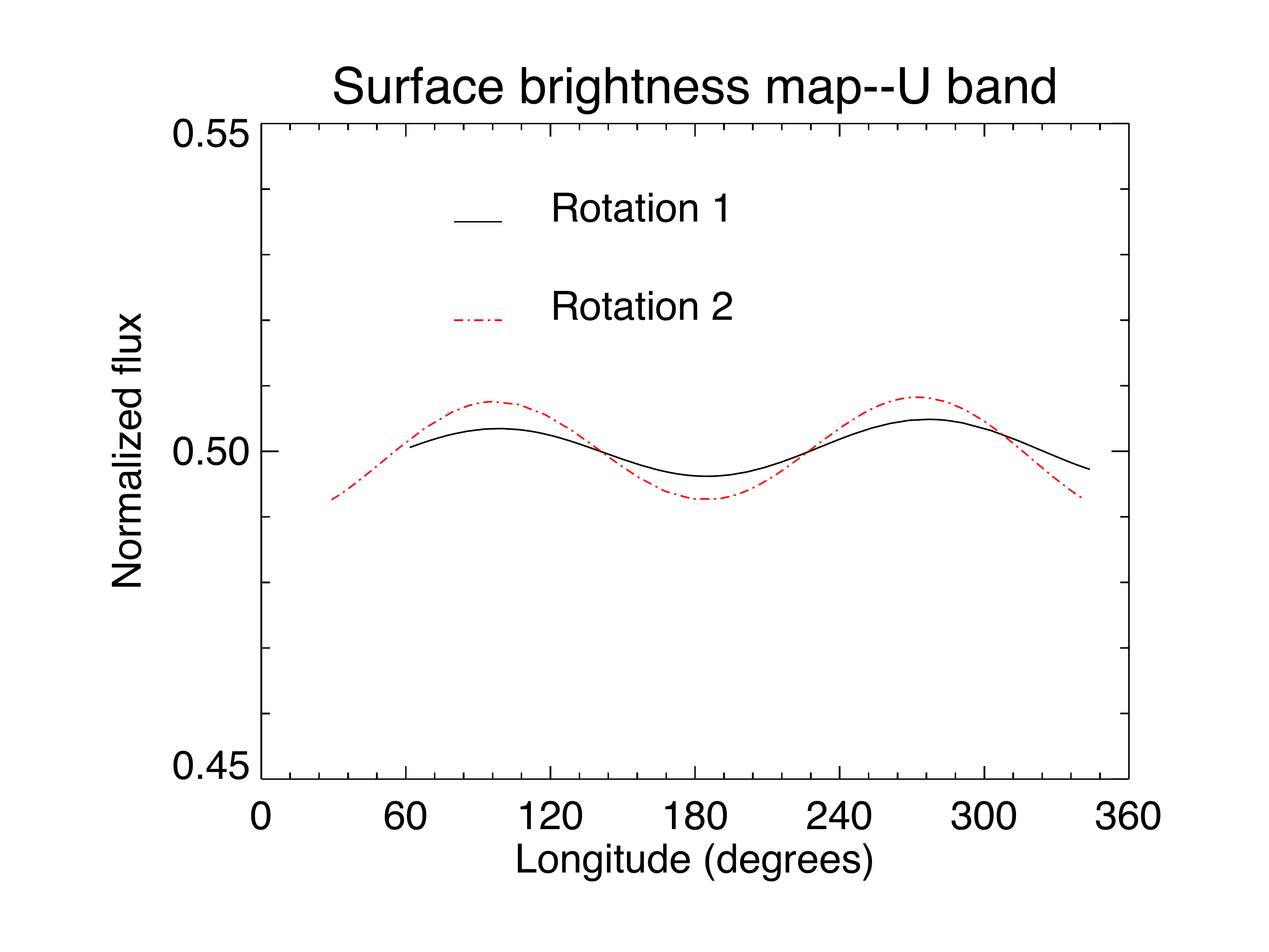}
\centering
\includegraphics[height=60mm]{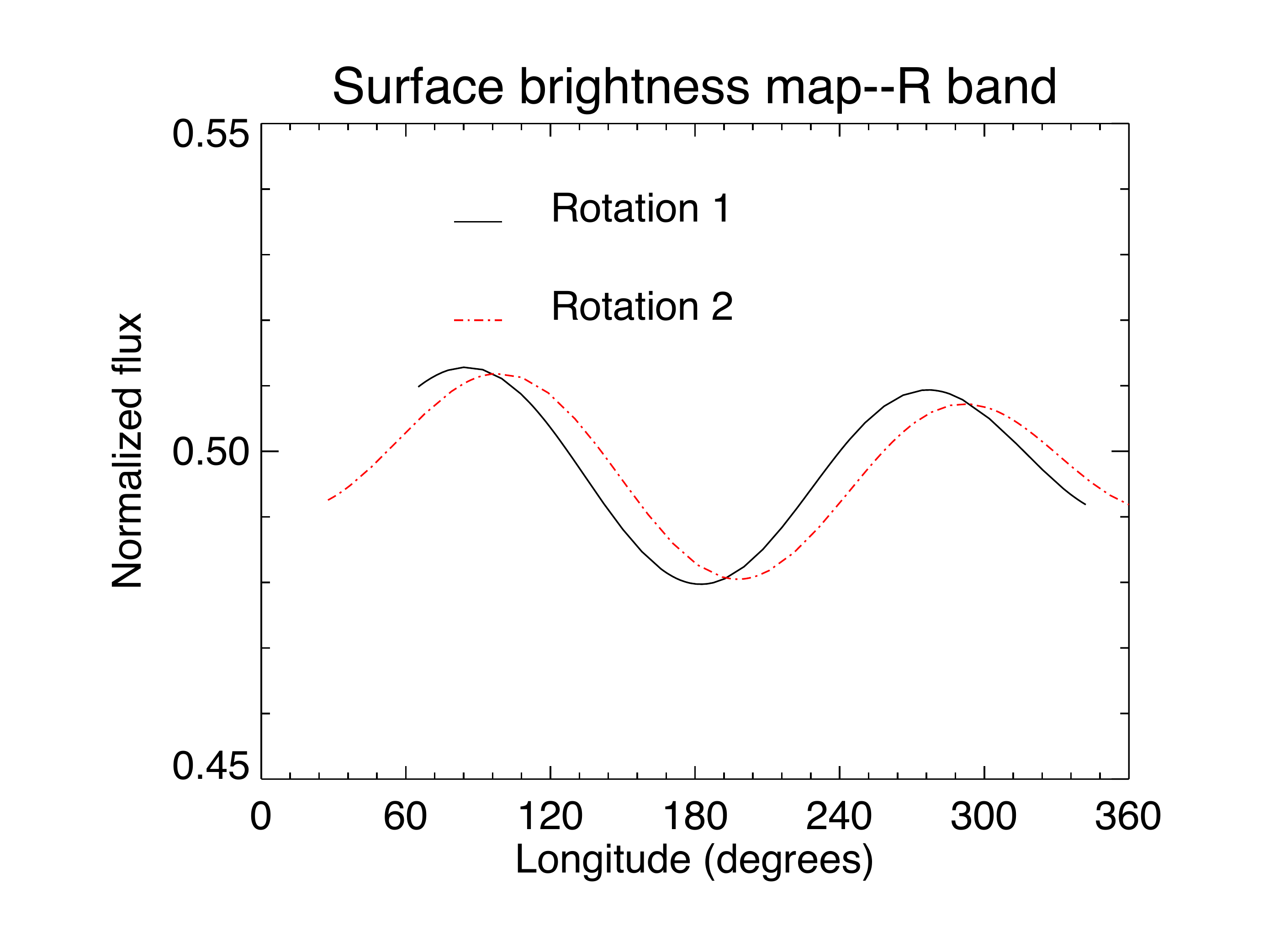}
\caption{Fourier surface brightness map of Jupiter based on its U (top) and 
R (bottom) bands, based on the first (black, solid lines) and the second 
(red, dashed--dotted lines) rotation.}
\label{fig:fft_map_ct_jup}
\end{figure}

\begin{figure}
\centering
\includegraphics[height=60mm]{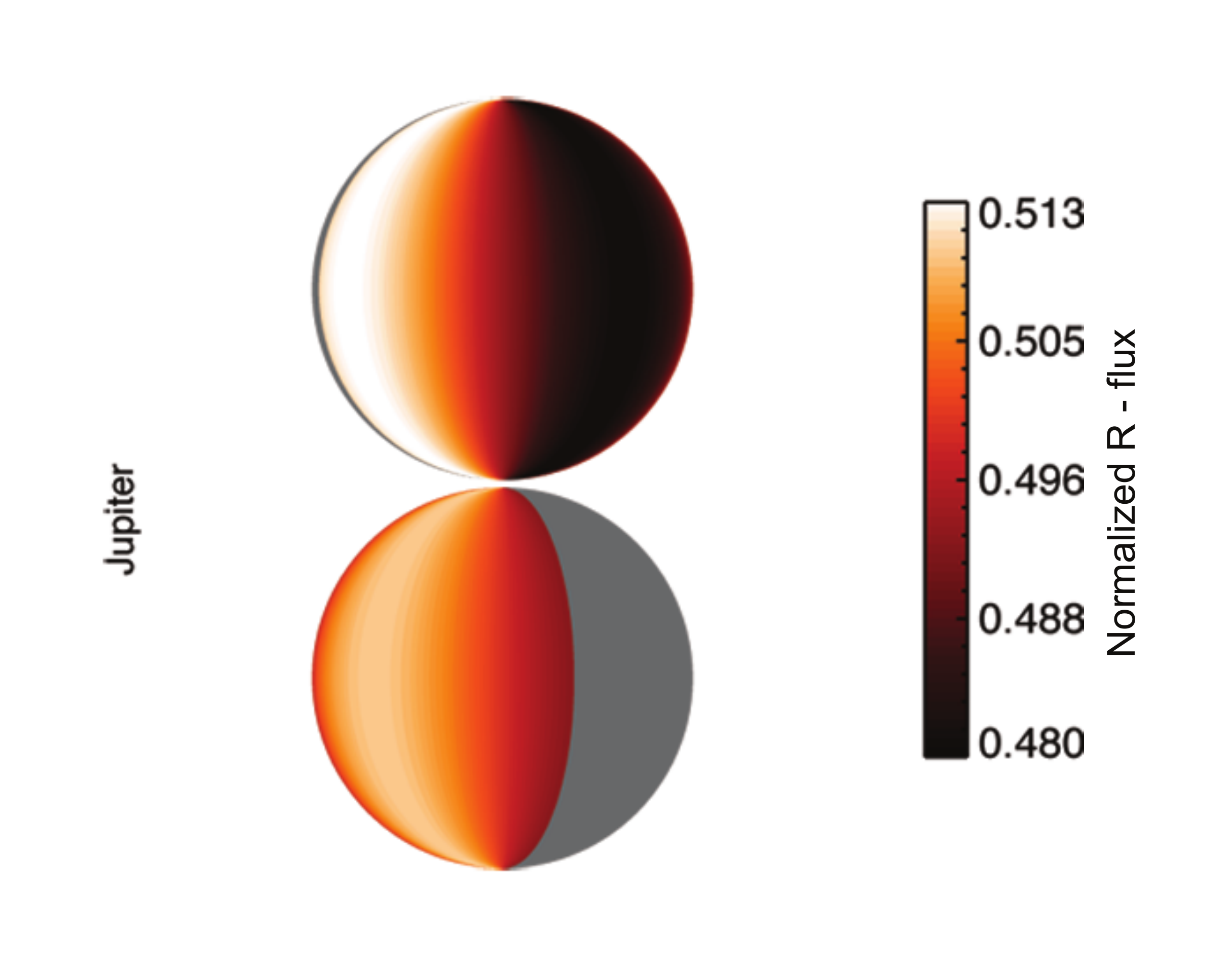}
\caption{Fourier surface brightness map of Jupiter based on its R--band, first rotation.
Upper panel: map centered at a longitude of $140^\circ$. Lower panel: map centered 
at a longitude of $320^\circ$.}
\label{fig:fft_map_r_jup}
\end{figure}

Fig.~\ref{fig:fft_map_u_jup} to~\ref{fig:fft_map_r_jup} show the 
longitudinal brightness maps of Jupiter in the U-- and R--bands. 
As discussed in Sect.~\ref{sect:map_jup}, we split the light curve for the two 
rotations and mapped each one separately. Fig.~\ref{fig:fft_map_u_jup} 
and~\ref{fig:fft_map_r_jup} show the map of the first rotation and, for comparison, 
Fig.~\ref{fig:fft_map_ct_jup} shows the map of both rotations 
in the U (top panel) and R (bottom panel) bands.
We ignored the first four snapshots of Jupiter due to an Europa intrusion, 
resulting in the first rotation maps (black lines of Fig.~\ref{fig:fft_map_ct_jup}) 
starting at $\sim$40 (rotational phase angle of $\sim$0.1).

In the U--band, features appearing in the map of the first rotation 
appear in the map of the second rotation as well, albeit 
with a different longitudinal size. 
The retrieved intensities of features in the second rotation 
are slightly higher than those of the first rotation. This is due to a slight 
increase ($\sim$0.03\%) in the normalized flux of the second rotation in comparison 
to the first rotation (see Fig.~\ref{fig:jup_lightcurves}).
In the R--band the maps of the two rotations are of equal brightness, 
but slightly offset.

Comparing the U--band maps with Jupiter HST images, 
we notice that the dark area around $\sim$340$^\circ$ of longitude 
coincides with the location of the GRS and Oval BA on the disk. In the 
first rotation map, the dark region appears broad and incorporates longitudes 
coexisting with the GRS and Oval BA on the HST snapshots. In the second 
rotation, the dark region appears at longitudes $\lesssim60^\circ$ and 
$\gtrsim340^\circ$, incorporating longitudes coexisting with the GRS and 
Oval BA on the HST snapshots.
The brightening around a longitude of $250^\circ$ could be 
related to a white plume appearing in the jovian disk. White plumes are thought 
to be the result of upwelling NH$_3$ clouds that freeze, resulting in high altitude 
fresh ice cloud \citep[][]{simonmiller01}. Finally, the darker area around 
$180^\circ$ corresponds to a featureless jovian disk.

Comparing the R--band maps with the HST images, we notice that 
the brighter area around a longitude of $\sim$100$^\circ$ corresponds to 
the snapshots in which the big hot spot is visible. The  
darker area around a longitude of 200$^\circ$ corresponds to snapshots 
in which smaller hot spots are visible on the disk. 
Finally, the brightening of the disk around 300$^\circ$ 
corresponds to images where the GRS appears on the disk (remember 
that as mentioned in Sect.~\ref{sect:phenom_jup} the GRS cannot 
be seen in the R--band, but appear as areas of equal brightness 
to the southern temperate belt).

 \subsection{A modified Jupiter} \label{sect:mod_jup}
 
To test \textit{Aeolus}, we simulated a Jupiter--like planet with 
extra spots at various locations and various sizes and contrast ratios 
and retrieved the maps of these ``modified'' Jupiters. 
In Table~\ref{table:jup_tsts} we summarize the various spot locations used.

\begin{table}[t]
\caption{Test cases for  \textit{Aeolus}. To test the sensitivity of \textit{Aeolus} we simulated a 
modified Jupiter with two spots at various longitudinal (l) and 
latitudinal ($\phi$) distances between them. The spot size and contrast ratio to the 
background TOA was kept constant for every set of simulations.}
\centering
\resizebox{.5\textwidth}{!} {
\begin{tabular}{c c c c c}
\hline
\hline
Test case & l$_1$(deg) &   l$_2$(deg) & $\phi_1$(deg) & $\phi_2$(deg)\\
\hline
1 &  130 & [145,149,154,164,174,184,204,224] & 0  & 0 \\
2 & 130 & 280 & 0 & [0,30,60]\\
3 & 130 & 130 & 0 & [20,30,50,80] \\
4 & 343 & N/A & [0,30,50,60,80] & N/A \\
\hline
\end{tabular}
}

\label{table:jup_tsts}
\end{table}

\begin{figure}
\centering
\includegraphics[height=124mm]{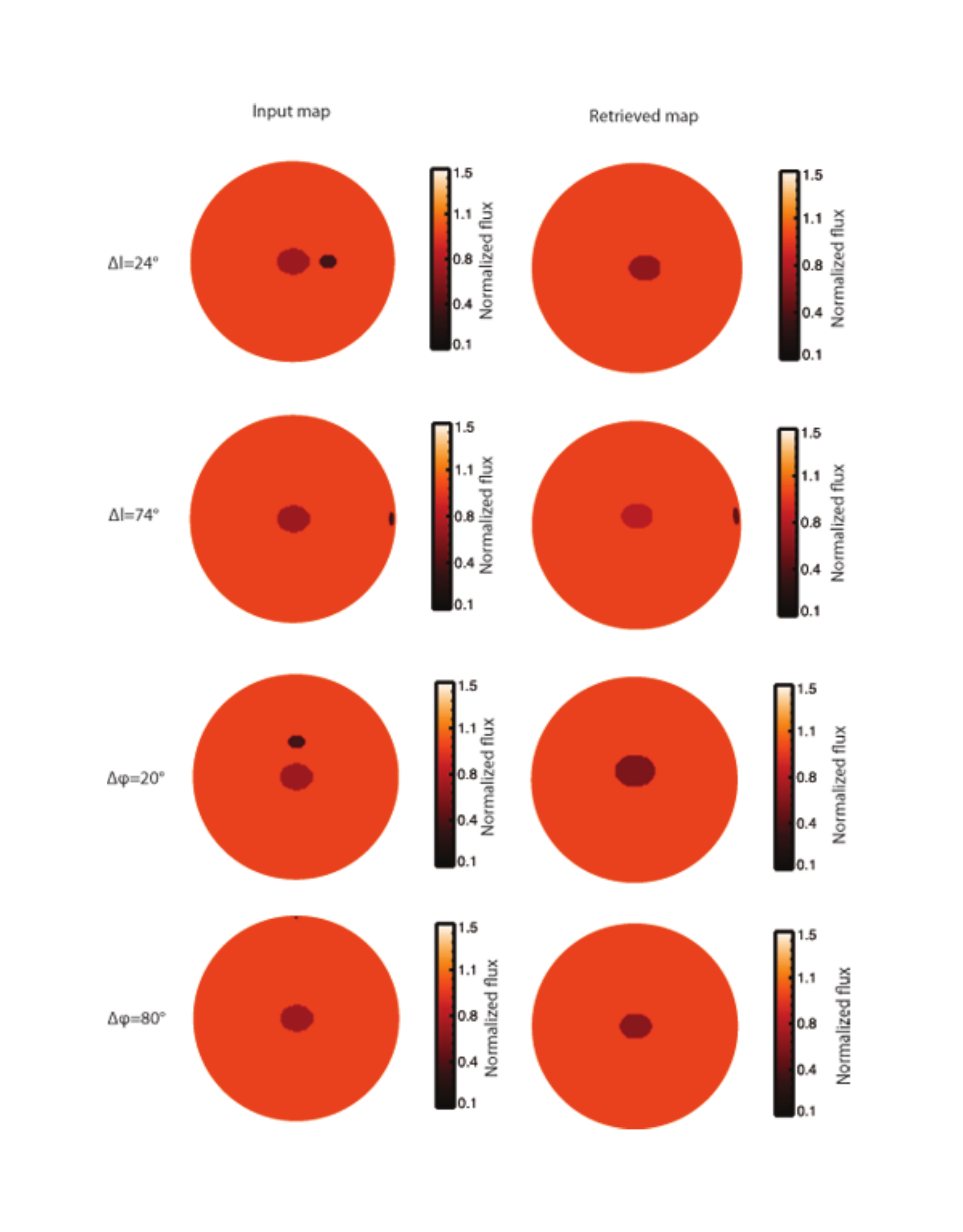}
\caption{Sample of input (left column) and \textit{Aeolus} retrieved maps (right columns), 
for test cases:1c (first row), 1g (second row), 3a (third row) and 3d (fourth row). For clarity we show 
the maps centered at 130$^\circ$ longitude.}
\label{fig:jup_combi}
\end{figure}

Initially, we simulated an atmosphere with two spots located at the equator and 
varied the longitudinal distance between them (see Table~\ref{table:jup_tsts}), to 
study the longitudinal sensitivity of our mapping code. We placed one spot 
at $l=130^\circ$, with a size of $s=18^\circ$ and a contrast ratio of 0.7, while the second 
spot had a size of $10^\circ$ and a contrast ratio of 0.4. In Table~\ref{table:jup_results} we 
show the number of spots retrieved from \textit{Aeolus}, its corresponding BIC, and whether the 
retrieved properties are (within the error bars) in agreement with the input properties or averaged 
between the two spot properties. For longitudinal spot separations (center to center) up to $34^\circ$, 
\textit{Aeolus} retrieved 1 spot with average properties, while for larger separations, it retrieved 
2 spots with properties that agreed, within the error bars, with the input properties. 
As an example, Fig.~\ref{fig:jup_combi} (upper half) shows the input map 
(left column) and the corresponding \textit{Aeolus} retrieved map (right column) for 
test cases 1c (first row) and 1g (second row). For clarity, we plot the maps centered at 
130$^\circ$ longitude.

We then placed the second spot at a longitude of $280^\circ$ and 
varied its latitude (see Table~\ref{table:jup_tsts}). We set the spot size equal to $20^\circ$ 
and contrast ratio to the background to $0.4$. \textit{Aeolus} retrieved 2 spots, whose 
longitude and size were, within the error bars, in agreement with the 
input properties (see Table~\ref{table:jup_results}). 
The latitudinal location and contrast ratio of the spots were retrieved slightly offset from 
the input values. 

We then modeled an atmosphere with one spot at a longitude of 130$^\circ$, a 
latitude of $0^\circ$, a size of $s=18^\circ$, and a contrast ratio of 0.7; and a second 
spot at the same longitude (130$^\circ$), and varied its latitude (see Table~\ref{table:jup_tsts}). 
We set the second spot's size to 10$^\circ$ and 
contrast ratio to the background at 0.4. \textit{Aeolus} retrieved 1 spot with all 
properties averaged (Table~\ref{table:jup_results}). As an example, Fig.~\ref{fig:jup_combi} 
(bottom half) shows the input map (left column) and the corresponding 
\textit{Aeolus} retrieved maps (right column) for test cases 3a (third row), and 3d (fourth row).
We note that the closer the second spot 
was to the pole, the closer the retrieved properties were to the equatorial spot's properties. 
We observed a similar behavior when mapping Jupiter based on its U--band light curve 
(Oval BA cannot be retrieved). This is due to a degeneracy among models 
with spots at different latitudes and with different contrast ratios/sizes when flux (without polarization) 
measurements are taken into account. We will discuss this problem further in 
Sect~\ref{sect:discussion}.

\begin{figure}
\centering
\includegraphics[height=60mm]{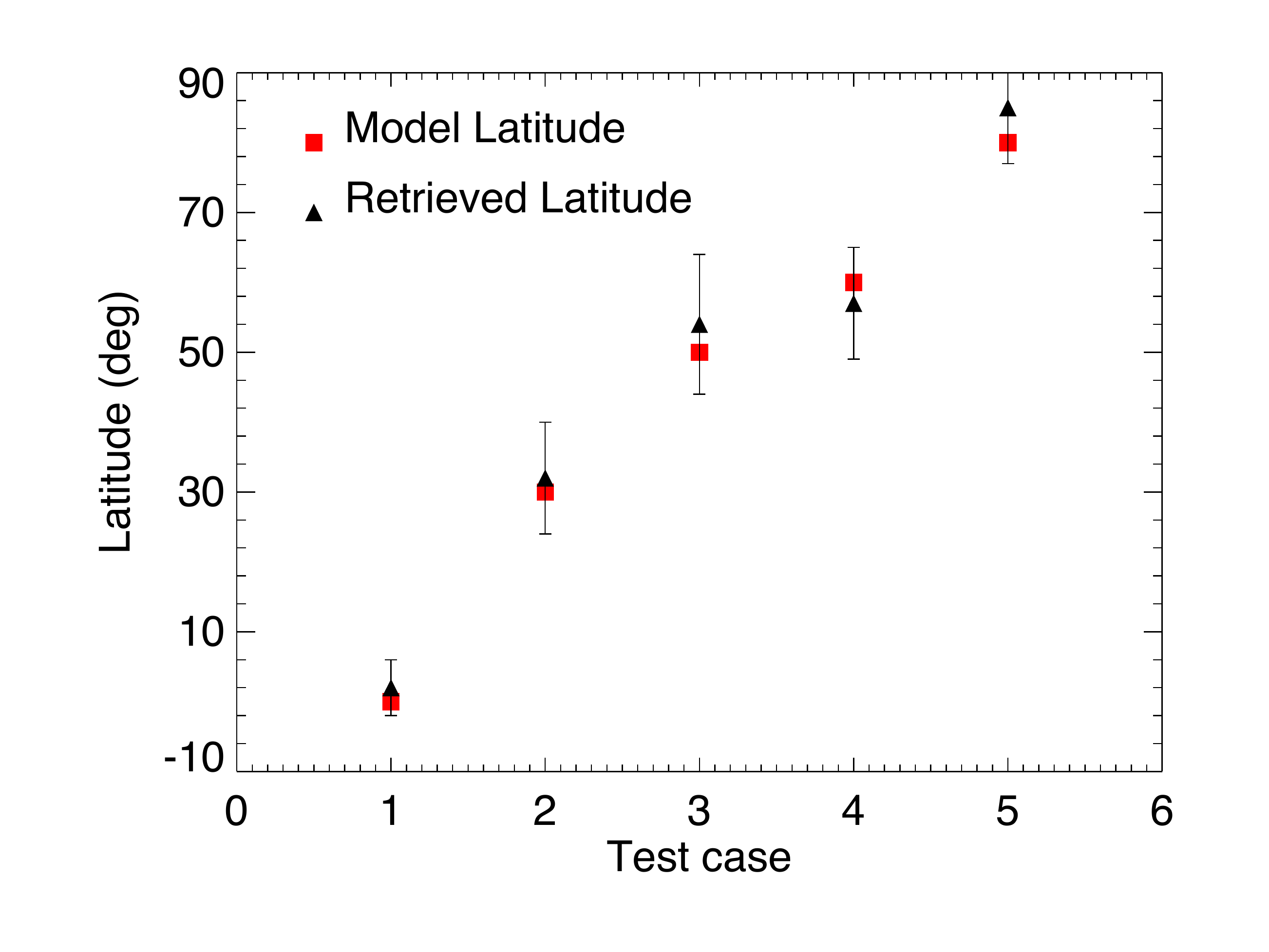}
\caption{Latitude of the spot of five model atmospheres (red squares), and corresponding 
\textit{Aeolus} retrieved latitudes (black triangles), with error bars. In high quality data \textit{Aeolus} 
can correctly identify the latitude of the elliptical features.}
\label{fig:inc_fit}
\end{figure}

We finally modeled an atmosphere with one spot, at a longitude of $343^\circ$, 
with a size of $s=27^\circ$ and a contrast ratio of 0.87, and varied its latitude through the 
following values: $0^\circ$, $30^\circ$, $50^\circ$, $60^\circ$ and $80^\circ$. Fig.~\ref{fig:inc_fit} shows the latitude of 
the spot for the five test cases (red squares), and the corresponding latitudes \textit{Aeolus} 
retrieved (black triangles), with error bars. \textit{Aeolus} retrieved the variation of the spot's latitude 
between our test cases, demonstrating the two--dimensionality of \textit{Aeolus} maps. 
We also tested the effect that an error in the estimated rotational period has on the 
retrieved maps. We varied the estimated rotational period by up to 10\% and compared the maps \textit{Aeolus} 
retrieved with those retrieved when the rotational period is known accurately. We found that the  
retrieved maps are in agreement ($\Delta l_{max}\sim$0.49\%, $\Delta\phi_{max}\sim$0.78\%, 
$\Delta s_{max}\sim$ 0.66\%, $\Delta f_{max}\sim$1.19\%), indicating that small uncertainties 
in the rotational period do not have a major impact on \textit{Aeolus} maps.

\begin{table}[t]
\caption{\textit{Aeolus} results for test cases of Table~\ref{table:jup_tsts}.}
\centering
\resizebox{.5\textwidth}{!} {
\begin{tabular}{c c c c c}
\hline
\hline
Test case & \# spots & BIC & Retrieved properties \\
\hline
1 (a) to 1 (d) & 1 & 16.4 & averaged \\ 
1 (e) to 1 (h) & 2 & 15.4 to 17.04 & in agreement \\
2 & 2 & 16.5 & in agreement \\
3 & 1 &  15.5 to 16.4 & averaged\\
4 & 1 & 16.2 to 20. & in agreement \\
\hline
\end{tabular}
}

\label{table:jup_results}
\end{table}

 \section{Brown dwarfs} \label{sect:bds}

Temporal variations in brown dwarf brightnesses indicate that their atmospheres 
present complex cloud structures \citep[][]{apai13}. 
Here we applied \textit{Aeolus} to map two rotating brown dwarfs in the  L/T transition, 
2MASSJ0136565+093347 (hereafter SIMP0136) and 
2MASSJ21392676+0220226 (hereafter 2M2139). We used observations 
that were taken by \citet[][]{apai13} using the G141 grism of the Wide Field Camera 3 on the 
Hubble Space Telescope (Project 12314, PI: Apai). These observations provide spatially 
and spectrally resolved maps of the variable cloud structures of these brown dwarfs.
For a detailed description of the data acquisition and reduction, 
we refer the reader to \citet[][]{apai13}. \citet[][]{apai13} performed synthetic 
photometry in the core of the standard J-- and H--
bands. 

\begin{figure}
\centering
\includegraphics[height=60mm]{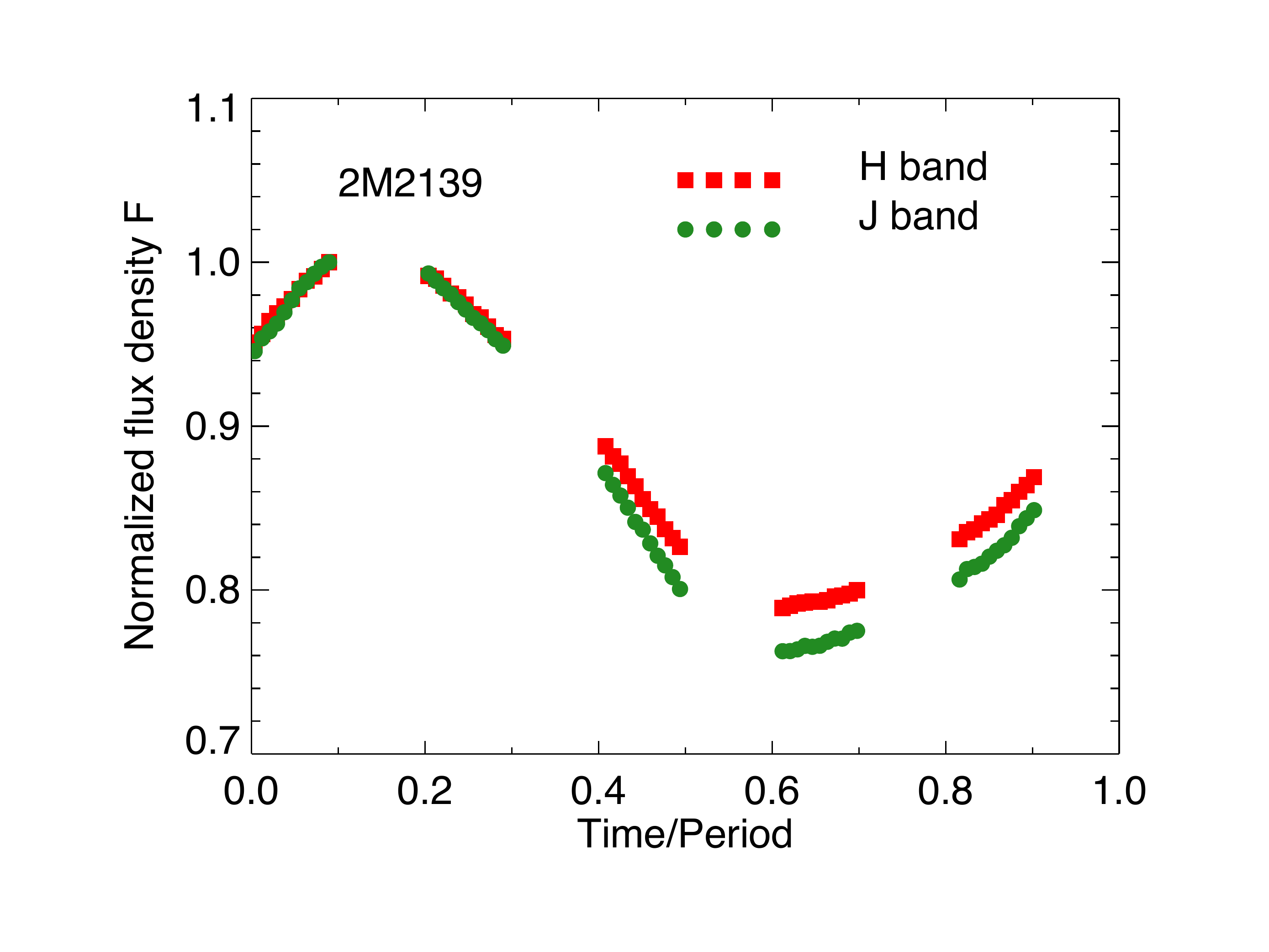}
\centering
\includegraphics[height=60mm]{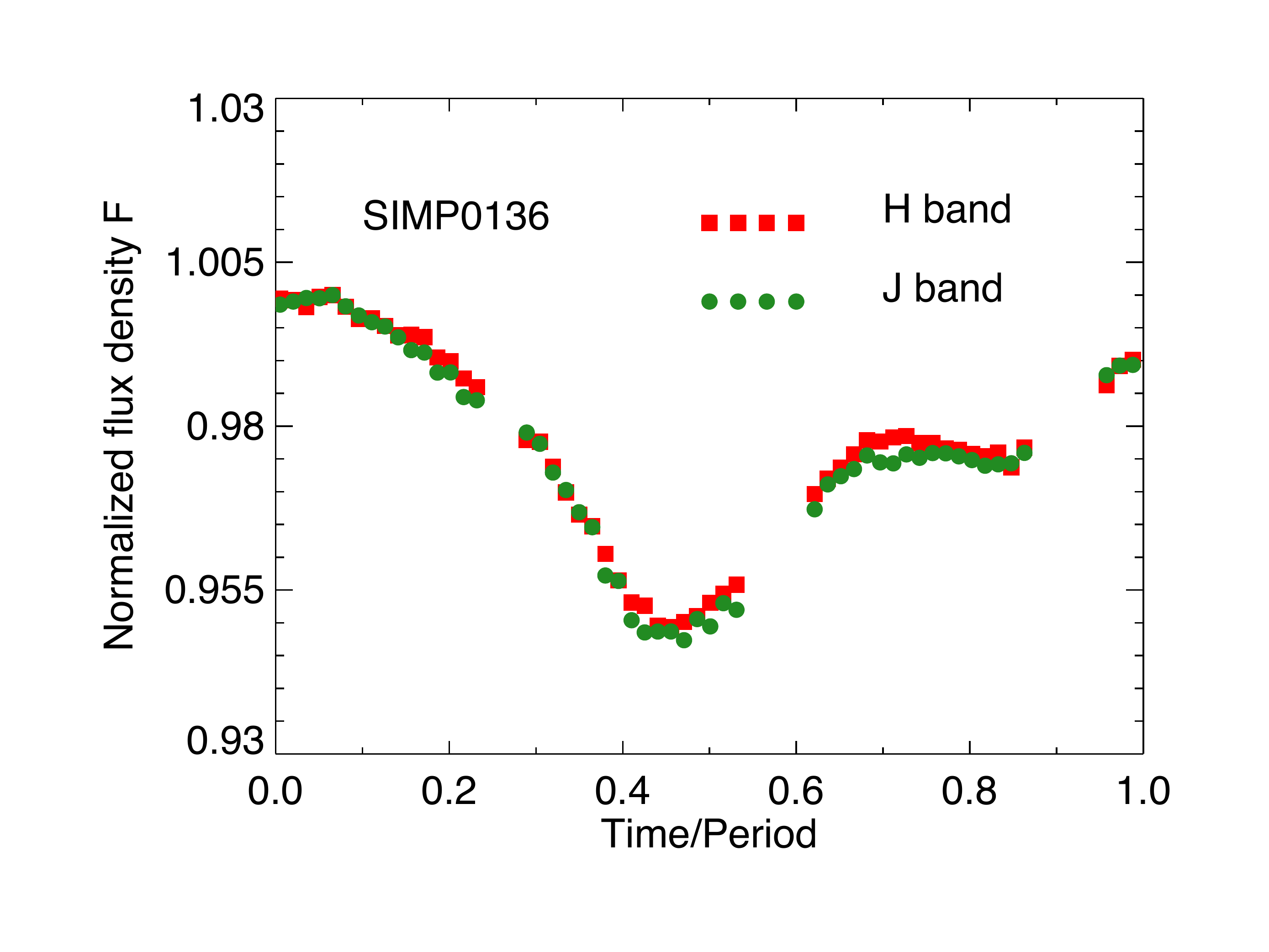}
\caption{Period folded H(red blocks), and J (green circles) light curves of 2MASSJ21392676+0220226 (top panel) 
and 2MASSJ0136565+093347 (lower panel).}
\label{fig:bd_lc}
\end{figure}

In Fig.~\ref{fig:bd_lc} we show the period--folded H (red blocks) and J (green circles) 
light curves of 2M2139 (top panel) and SIMP0136 (lower panel). 
Both 2M2139 and SIMP0136 exhibit brightness variations in 
the H-- and J--bands, with peak--to--peak amplitudes of 27\% and 4.5\% respectively. 
Both targets' light curves vary in a similar manner, independent of 
the observational wavelength. Given that as previously discussed (Sect.~\ref{sect:phenom_jup}), 
different wavelengths probe different pressure layers, the similar appearance 
of 2M2139 and SIMP0136 in the H-- and J--bands indicates a similar 
TOA map for the different pressure levels.

\subsection{2M2139}

2M2139 is classified as a T1.5 by \citet[][]{burgasser06} based on 
NIR observations. Later observations suggested 2M2139 could be 
a binary composite of an L8.5$\pm$0.7 and a 
T3.5$\pm$1.0 based on SpeX spectra \citep[][]{burgasser10}, even though a spectral modeling study by 
\citet[][]{radigan12} reached a different conclusion and 
high--resolution HST observations detected no evidence for a companion \citep[][]{apai13}.
Ground--based photometry of 2M2139 suggested light curve evolution on timescales of days, 
indicating a considerable evolution of cloud cover in its atmosphere \citep[][]{radigan12}. 
\citet[][]{radigan12} observed a very large variability of up to $26\%$ 
in the J--band and a period of 7.721$\pm$0.005 hr. 
\citet[][]{apai13} carried out time--resolved HST near--infrared spectroscopy that covered a 
complete rotational period. This dataset showed that rotational modulations are gray, i.e. only 
weakly wavelength--dependent. State--of--the--art radiative transfer modeling of the color--magnitude 
variations demonstrated that the changes are introduced by cloud thickness variations 
(warm thin and cool thick clouds). PCA analysis showed that $>99\%$ of the spectral variations 
can be explained with only a single principal component, arguing for a single type of cloud 
feature \citep[][]{apai13}. Light curve modeling found that 
three--or--more--spot models are needed to explain the observed light curve shapes.

\subsection{SIMP0136}

SIMP0136 is a T2.5 dwarf \citep[][]{artigau06}, with a period of 2.3895$\pm$0.0005 hr 
and exhibits peak to peak variability of up to 4.5$\%$ in the J-- and H-- 
bands \citep[][]{artigau09}. SIMP0136 shows a significant night-to-night evolution 
\citep[][]{artigau09,apai13,metchev13} even though it does not appear to be a binary 
\citep[][]{goldman08,apai13}. Time-resolved HST near-infrared spectroscopy by \citet[][]{apai13} found that 
the observed variations of SIMP0136 are nearly identical to those observed in 2M2139 and are also interpreted 
by a combination of thin clouds with large patches of cold and thick clouds.

\subsection{Comparison of \textit{Aeolus} with Fourier and PCA maps of 2M2139 and SIMP0136}

We applied \textit{Aeolus} to the light curves of Fig.~\ref{fig:bd_lc} and 
compared the retrieved maps of 2M2139 and SIMP0136 with 
the corresponding maps using Fourier decomposition and 
with \textit{Stratos} maps produced by \citet[][]{apai13}.

\begin{figure}
\centering
\includegraphics[height=55mm]{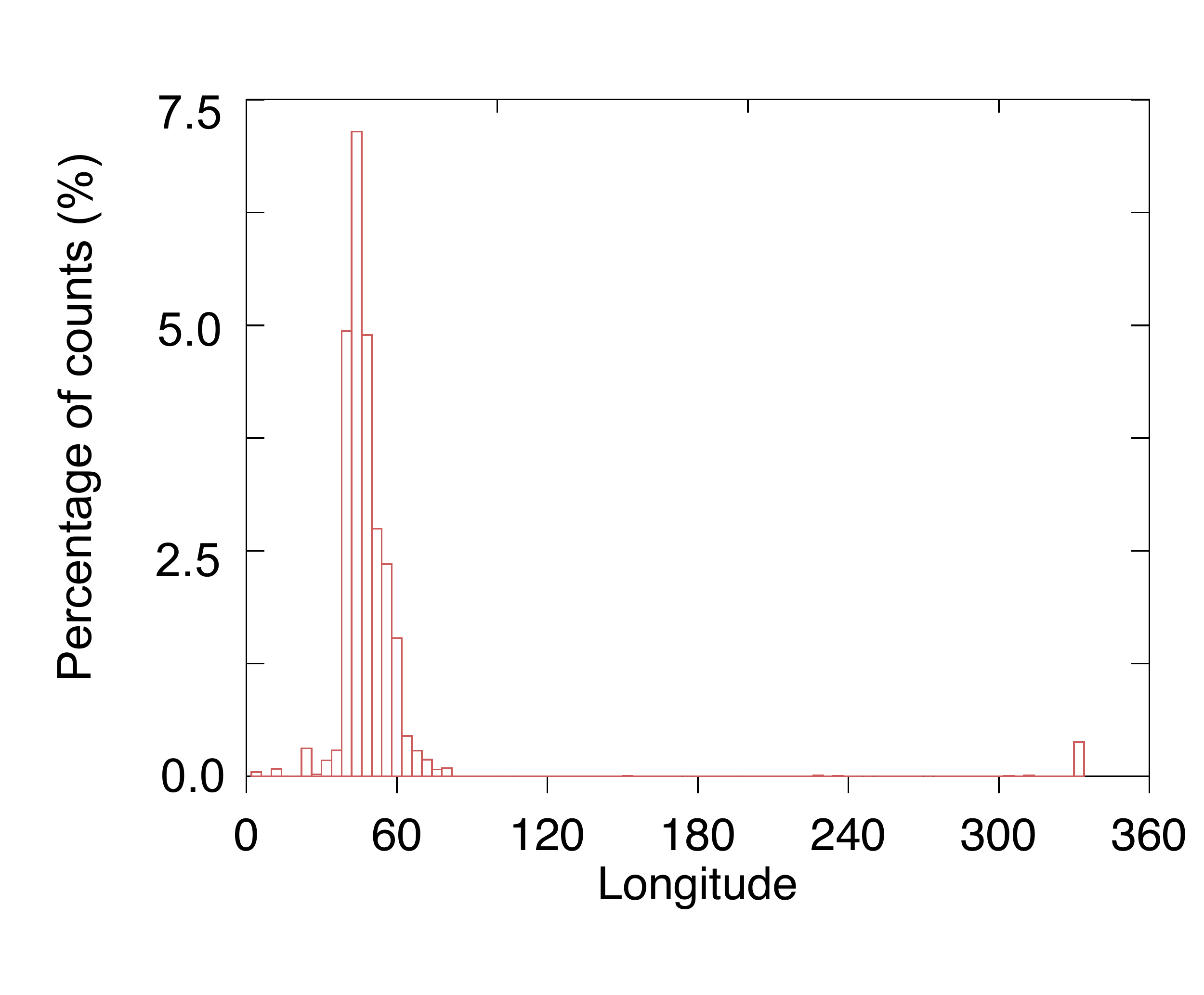}
\centering
\includegraphics[height=60mm]{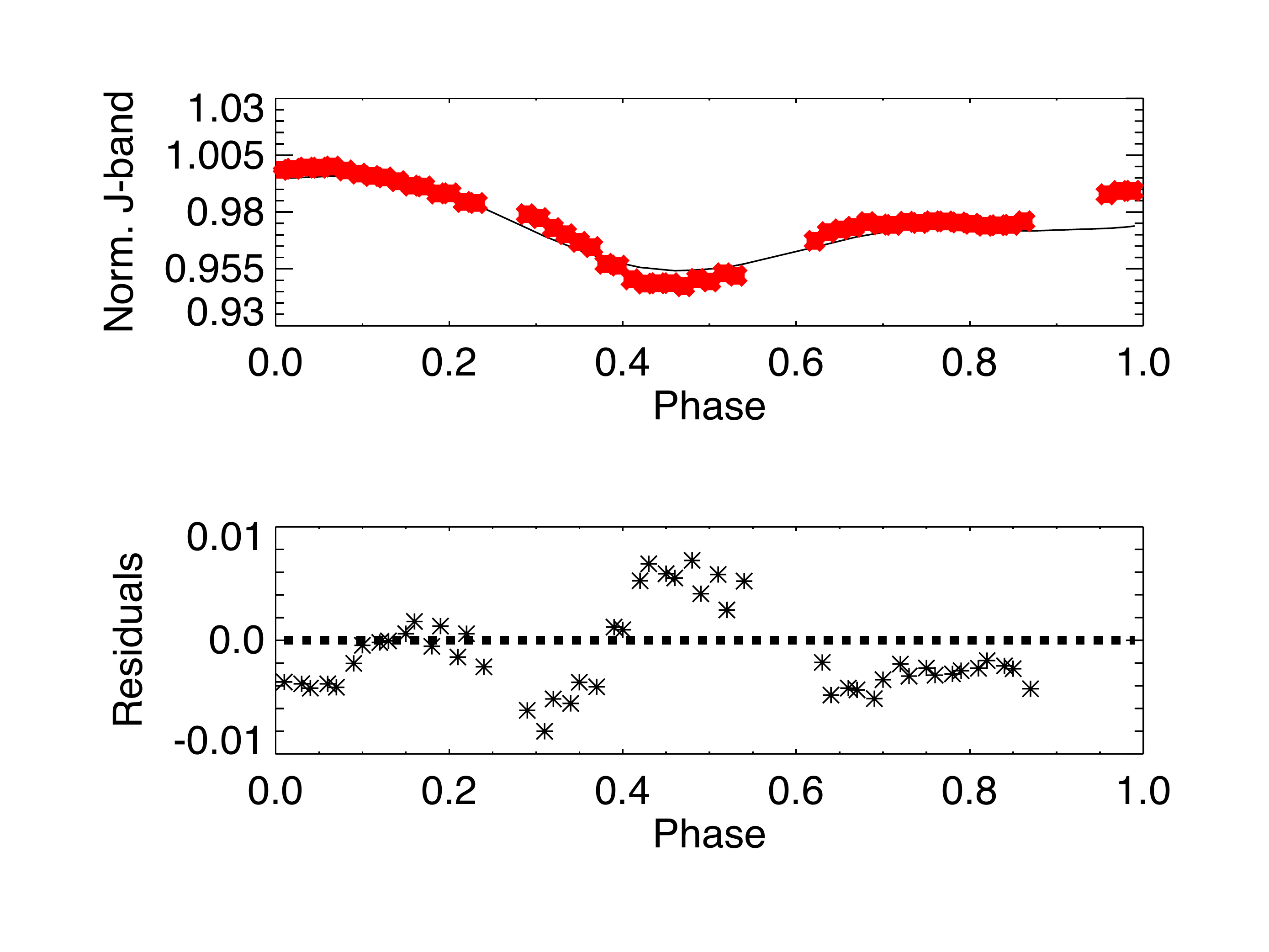}
\caption{Posterior distribution for the longitude of spot 1 of 2M2139 (top panel); 
normalized J--band light curve of 2M2139 (red triangles) with 
error bars and best-fit \textit{Aeolus} light curve (black, solid line) (middle panel); and 
residuals (bottom panel).}
\label{fig:post_distr_2m2139}
\end{figure}

\begin{figure}
\centering
\includegraphics[height=60mm]{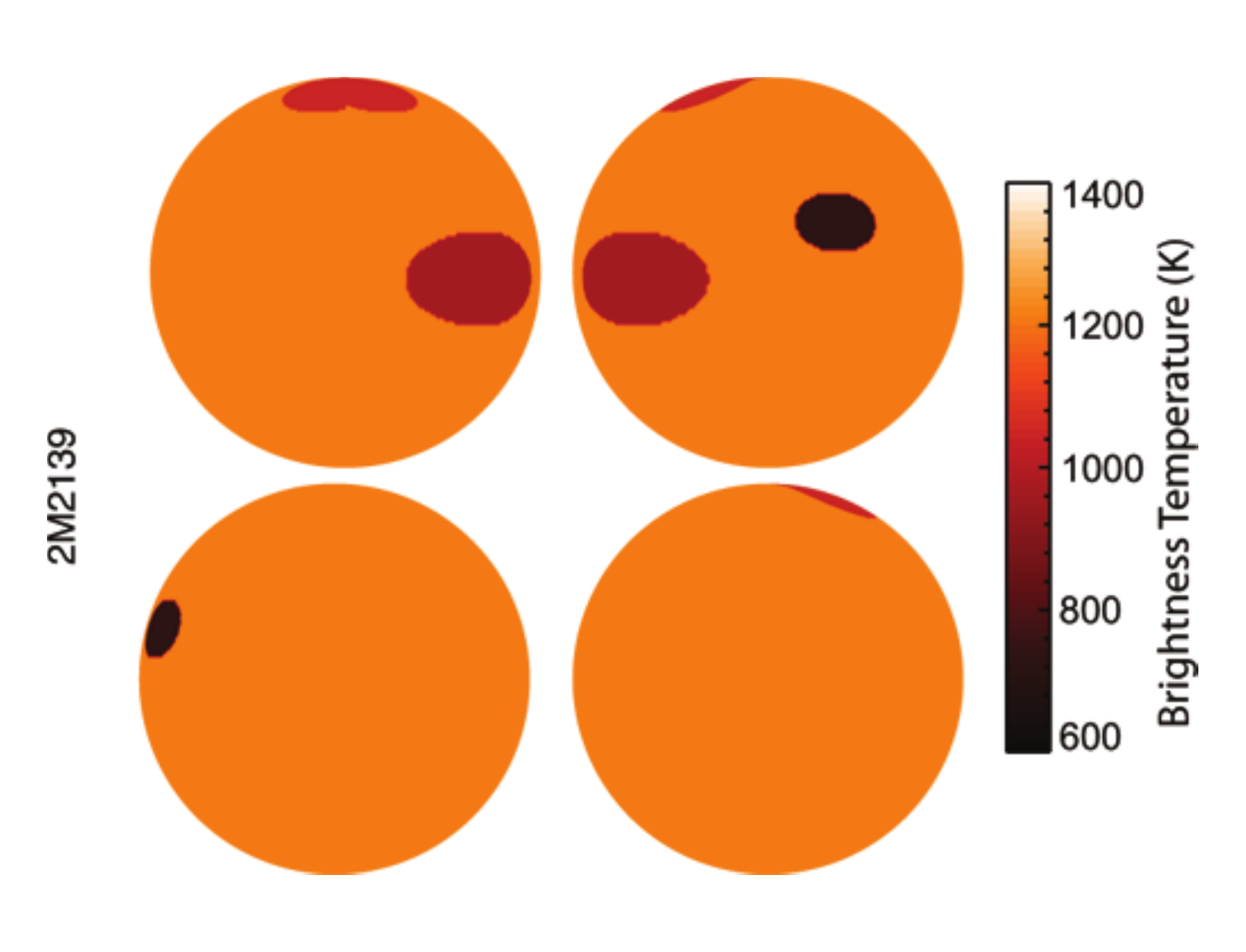}
\hspace{0.2cm}
\centering
\includegraphics[height=60mm]{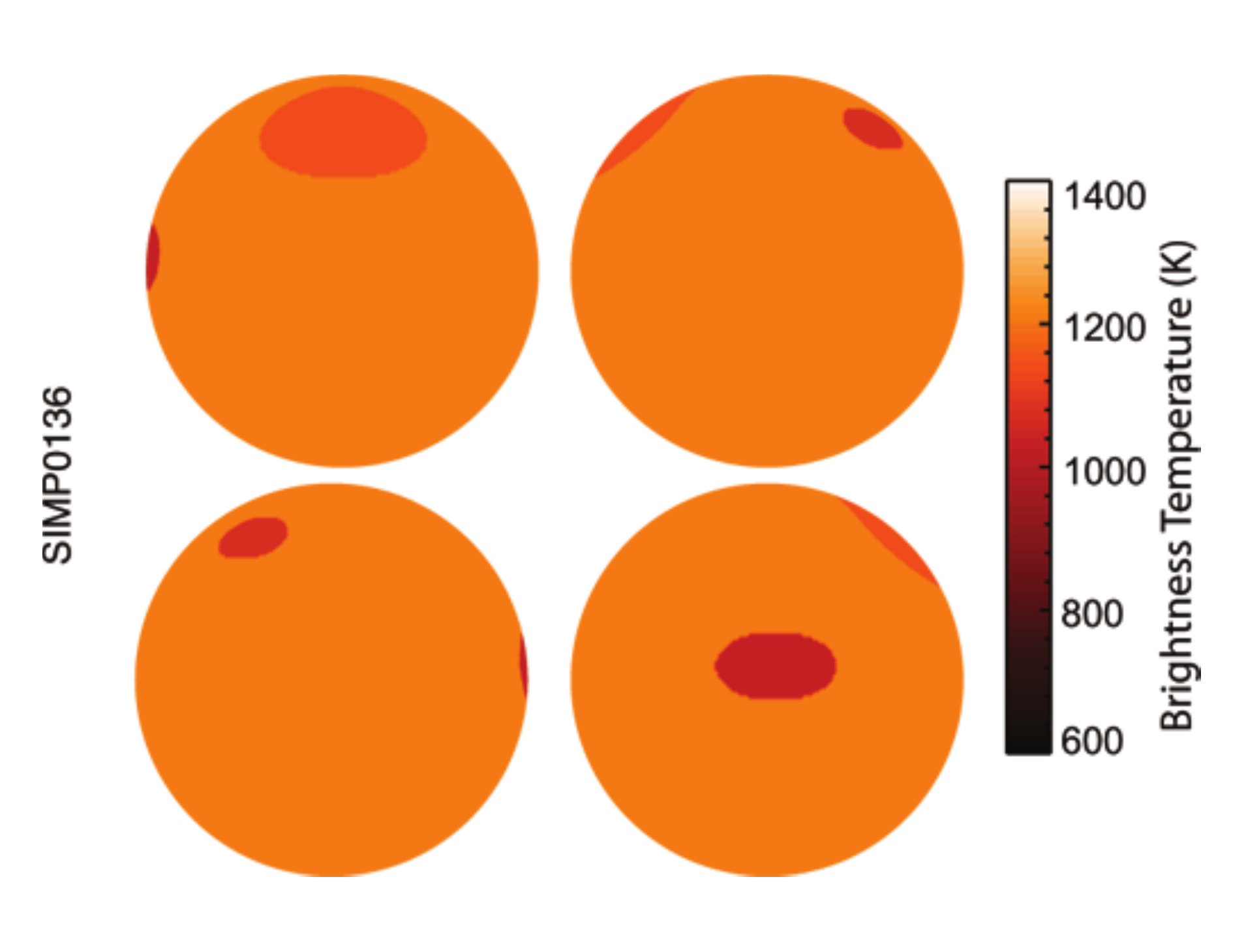}
\caption{2M2139 (top four maps) and SIMP0136 (bottom four maps) brightness temperature maps from applying 
\textit{Aeolus} on the J band light curves of Fig.~\ref{fig:bd_lc}. The maps are centered 
at $0^\circ$ of longitude (upper left map), $90^\circ$ of longitude (upper right map), 
$180^\circ$ of longitude (lower left map) and $270^\circ$ of longitude (lower right map). }
\label{fig:aeolus_map_bds}
\end{figure}

\begin{figure}
\centering
\includegraphics[height=60mm]{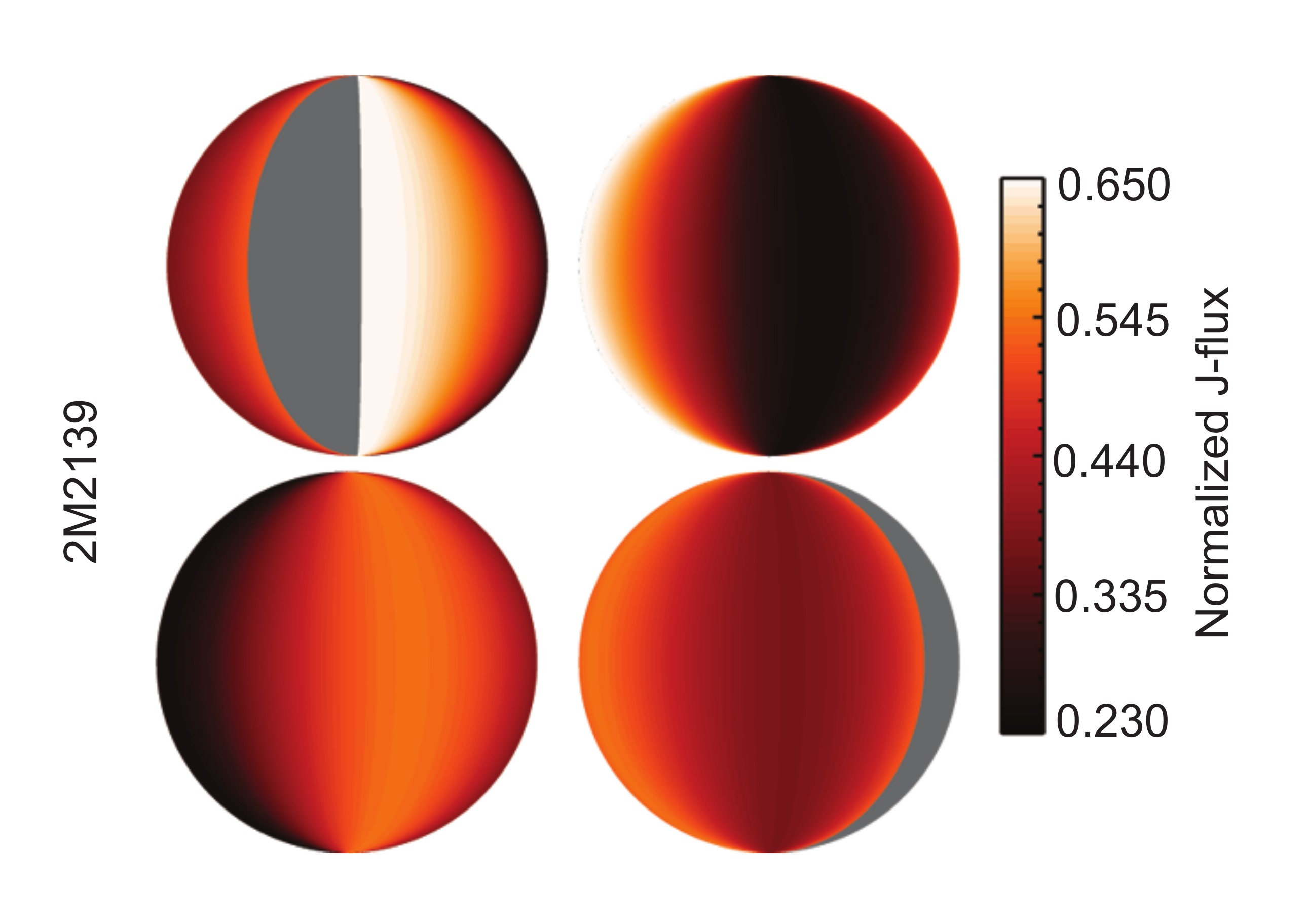}
\centering
\vspace{6pt}
\includegraphics[height=60mm]{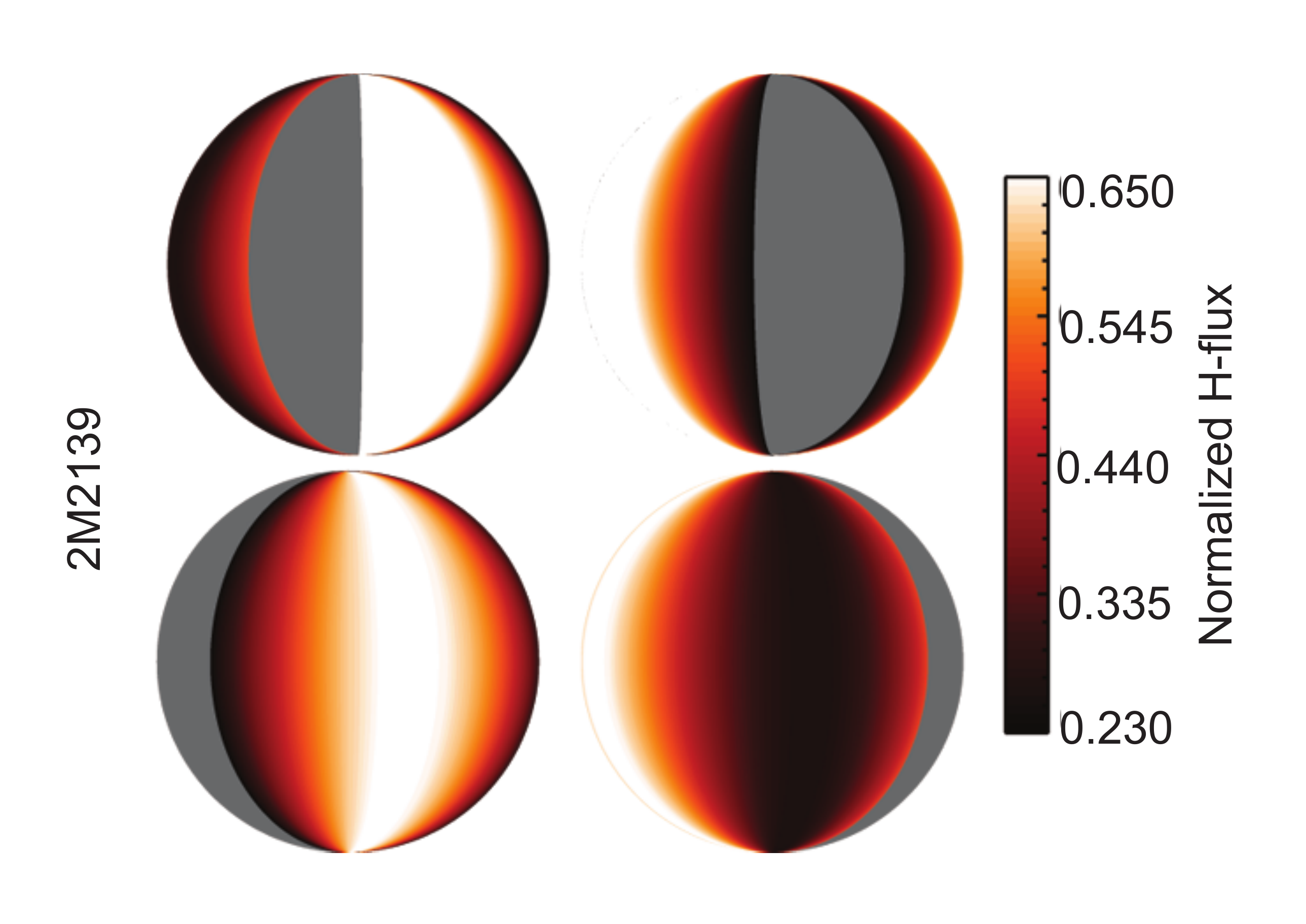}
\caption{2M2139 Fourier maps from the J (top four maps) and H (bottom four maps) band centered 
at $0^\circ$ (top left), $90^\circ$ (top right), $180^\circ$ (bottom left) and 
$270^\circ$ (bottom right) longitude. Dark grey indicates areas without data 
(due to the lack of data points above a rotational phase of 0.9). }
\label{fig:complete_2m2139_map}
\end{figure}

\begin{figure}
\centering
\includegraphics[height=58mm]{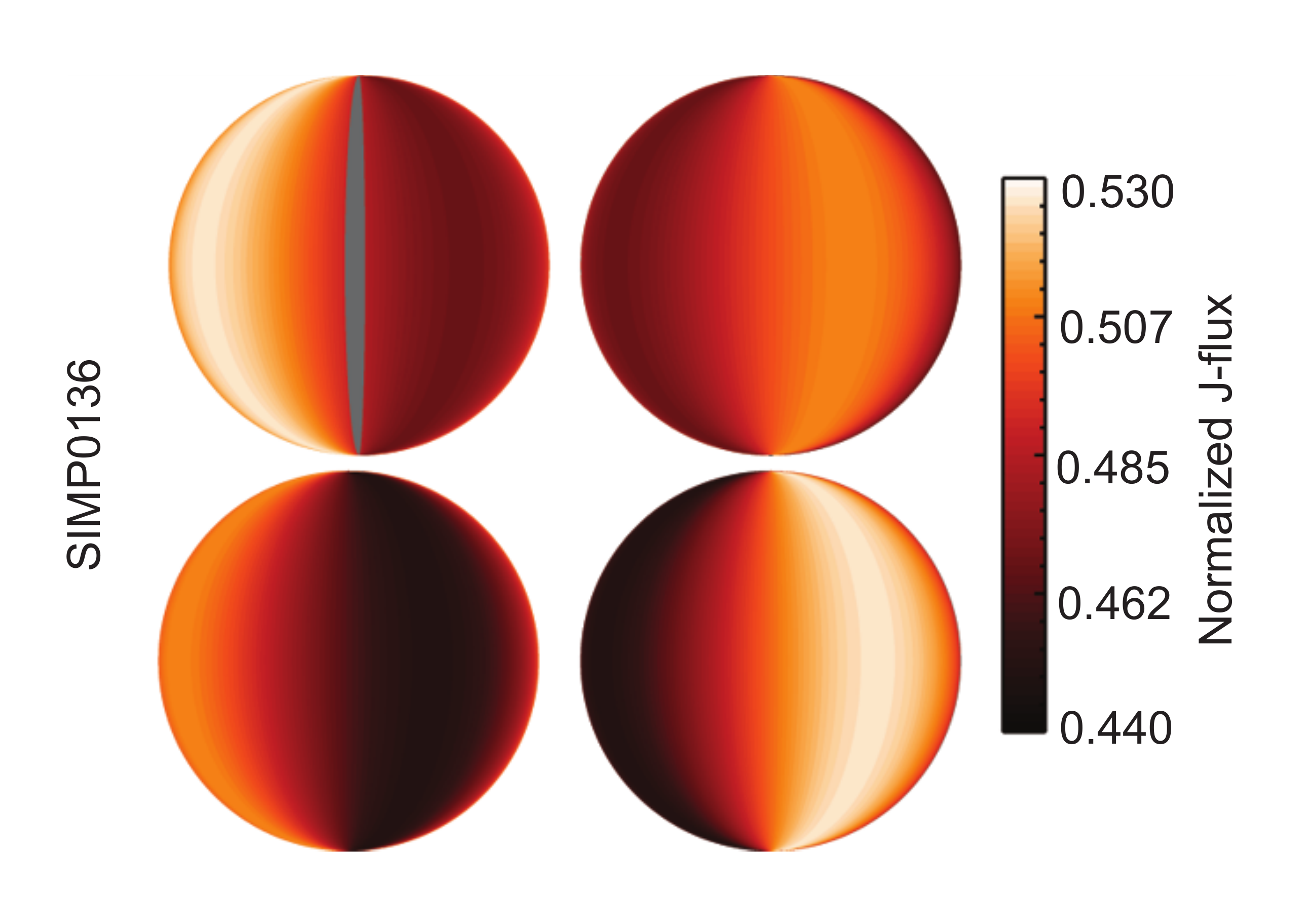}
\centering
\vspace{6pt}
\includegraphics[height=58mm]{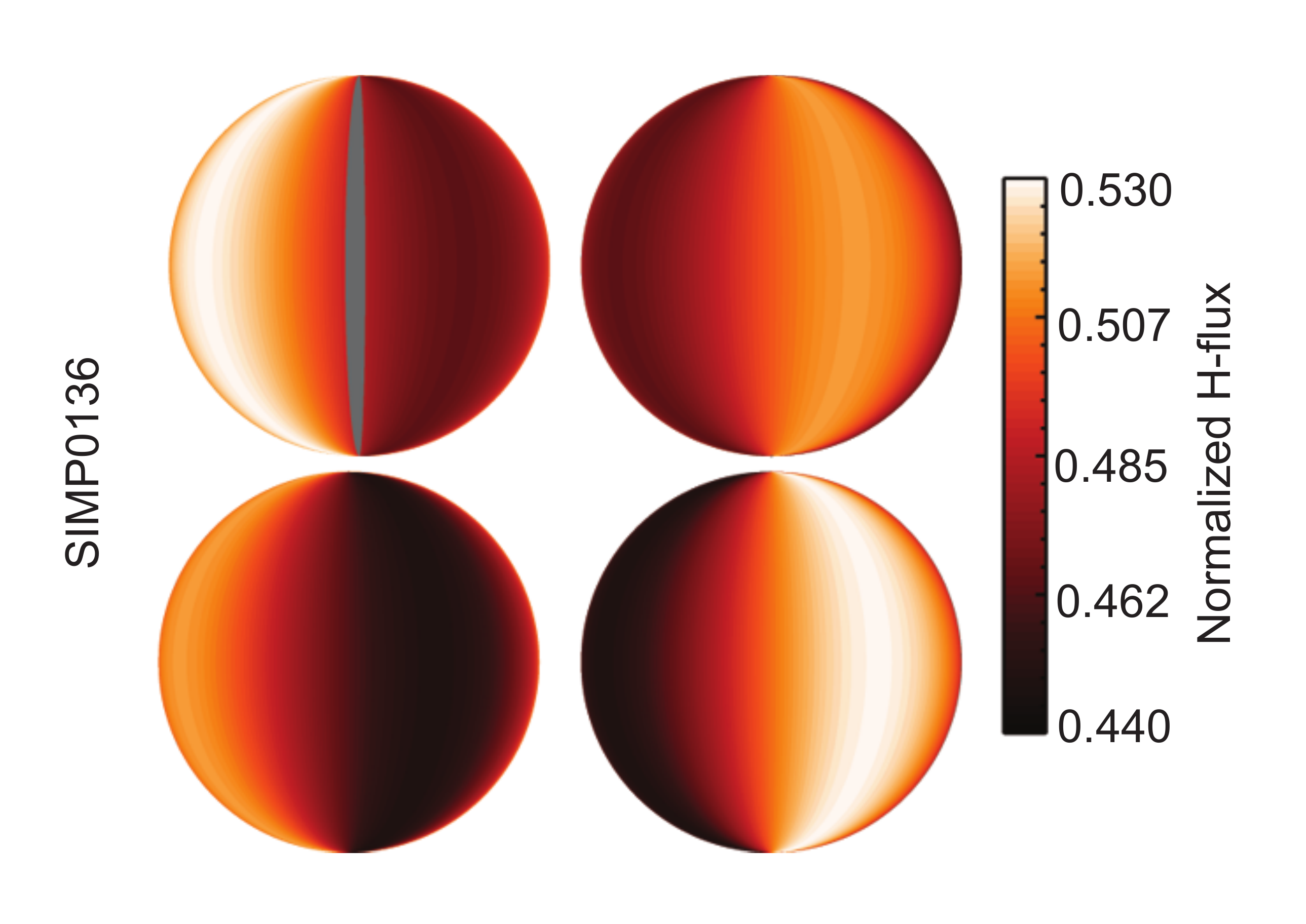}
\caption{Same as Fig.~\ref{fig:complete_2m2139_map} but for SIMP0136.}
\label{fig:complete_simp0136_map}
\end{figure}

Initially, we applied \textit{Aeolus} to the 2M2139 light curves of Fig.~\ref{fig:bd_lc}. 
Fig.~\ref{fig:post_distr_2m2139} shows the 
posterior distribution of the longitude of spot 1 (top panel); the normalized J--band light curve (red 
triangles) with error bars and best-fit \textit{Aeolus} light curve (black, solid line) (middle panel); and 
the corresponding residuals (bottom panel). 
Based on the J--band light curve, \textit{Aeolus} retrieved 3 spots (BIC 30) 
with (longitude, latitude) = ($111^\circ\pm15^\circ$, $15^\circ\pm10^\circ$), ($45^\circ\pm5^\circ$,  
$2^\circ\pm10^\circ$) and ($344^\circ\pm10^\circ$, $77^\circ\pm15^\circ$), with respective 
sizes of $13^\circ\pm3^\circ$, $27^\circ\pm4^\circ$ and $39^\circ\pm5^\circ$ and contrast ratios 
of $0.18\pm0.10$, $0.57\pm0.07$ and $0.79\pm0.04$. A similar map was retrieved 
based on the H--band light curve.

We then applied \textit{Aeolus} on the SIMP0136 light curves of Fig.~\ref{fig:bd_lc}. 
Based on the J--band light curve, \textit{Aeolus} retrieved 3 spots (BIC 51) with 
(longitude, latitude) = ($272^\circ\pm21^\circ$, $4^\circ\pm7^\circ$), ($143^\circ\pm20^\circ$, 
$47^\circ\pm17^\circ$) and ($0^\circ\pm5$, $49^\circ\pm15^\circ$), with respective 
sizes of $18.57^\circ\pm2.6^\circ$, $17^\circ\pm2^\circ$ and $37.5^\circ\pm1.8^\circ$ and contrast ratios 
of $0.77\pm0.07$, $0.87\pm0.18$ and $1.12\pm0.05$. A similar map was retrieved 
based on the H--band light curve.

In summary, \textit{Aeolus} found that both 2M2139 and SIMP0136 are covered by three spots, 
with a longitudinal coverage of 21\%$\pm$3\% and 20.3\%$\pm$1.5\% respectively 
(see Fig.~\ref{fig:aeolus_map_bds}).The size of the larger spot in 2M2139 was 
found to be $39^\circ\pm11^\circ$ and in SIMP0136 $37.5^\circ\pm1.8^\circ$. 2M2139's 
spots are darker than the background TOA, while SIMP0136 has two dark and one 
brighter than the background TOA spots. Assuming that brightness variations across the 
TOA are due to the different temperature of the areas observed, 
we can calculate the brightness temperature variations across the TOA. This would be, 
for example, the case when due to thinner clouds we see deeper, hotter layers of the atmosphere.
In Fig.~\ref{fig:aeolus_map_bds} we show 2M2139 and SIMP0136 brightness temperature maps, 
assuming the background TOA has a brightness temperature of 1100 K \citep[following][]{apai13}. 
The darkest spot of 2M2139 is $\sim380$ K cooler and its brightest spot is 
$\sim63$ K cooler than the background TOA. SIMP0136's darkest spot is $\sim70$ K cooler than 
the background TOA, while its brightest spot is $\sim32$ K hotter than the background TOA. 

We then applied the Fourier mapping technique to the 
light curves of Fig.~\ref{fig:bd_lc}. Figs.~\ref{fig:complete_2m2139_map} 
and~\ref{fig:complete_simp0136_map} show the maps of 2M2139 and SIMP0136 
respectively, in the J (top panel) and H (bottom panel) bands. 
As expected from the similarity of the light curves, the retrieved maps look similar 
in the two wavelengths. 

The J-band surface brightness map for 2M2139, relative to the global average, is bright for  
$280^\circ\lesssim l \lesssim 330^\circ$, and dark for $30^\circ\lesssim l \lesssim 100^\circ$ 
and $140^\circ\lesssim l \lesssim 230^\circ$. 
A brightening around $120^\circ$ corresponds to a bump in the light curve around a phase of 0.4.    
Given the amplitude of the flux increase (0.6\% with respect to a sinusoidal fit) and the 
uncertainty of  0.04\%, we conclude that this bump is due to an actual 
feature in the brown dwarf atmosphere. 2M2139's H--band map is similar to 
its J--band map but heterogeneous features appear less bright and narrower 
(by $\sim10^\circ$) than their J--band counterparts. These differences 
can be traced back to the differences in the H and J band light curves of Fig.~\ref{fig:bd_lc}.

The J-band surface brightness map for SIMP0136, relative to the global average, is bright for 
$40^\circ\lesssim l \lesssim 70^\circ$ and $220^\circ\lesssim l \lesssim 270^\circ$, and 
dark for $100^\circ\lesssim l \lesssim 200^\circ$ and $310^\circ\lesssim l \lesssim 340^\circ$. 
SIMP0136's H--band map is similar to its J--band map.

We could interpret our retrieved Fourier maps as finding two large scale 
heterogeneities on 2M2139 and three smaller 
scale heterogeneities on SIMP0136. In this scenario, 2M2139's heterogeneities 
have a longitudinal coverage of 50\% and SIMP0136's heterogeneities 
have a longitudinal coverage of 39\%. 

\citet[][]{apai13} using principal component analysis (PCA) and the 
mapping package \textit{Stratos}, found that only two kinds of 
clouds are necessary to describe the observed signals of 2M2139 and SIMP0136. 
One cloud is the ``background'' and the other needs to be distributed in 
\textit{at least} three spots. \citet[][]{apai13} found that the spots have 
a longitudinal coverage of 20\% to 30\% and that the diameter of the larger 
spot is $\sim$60$^\circ$.Finally, the spots need to have a brightness difference 
to the background by a factor of three.

\textit{Aeolus} agrees on the amount and 
longitudinal coverage of spots at the TOA of 2M2139 
and SIMP0136 with \textit{Stratos}, while Fourier mapping hints to potentially 
higher longitudinal coverage. The contrast ratios of spots \textit{Aeolus} 
retrieved on SIMP0136 agree within the error bars with the \citet[][]{apai13} 
results, while the 2M2139 darker spot is considerably darker. Finally, the maximum 
size of the spots retrieved by \textit{Aeolus} appears 
smaller than the maximum size found with \textit{Stratos}.

\section{Discussion}~\label{sect:discussion}

\begin{figure}
\centering
\includegraphics[height=58mm]{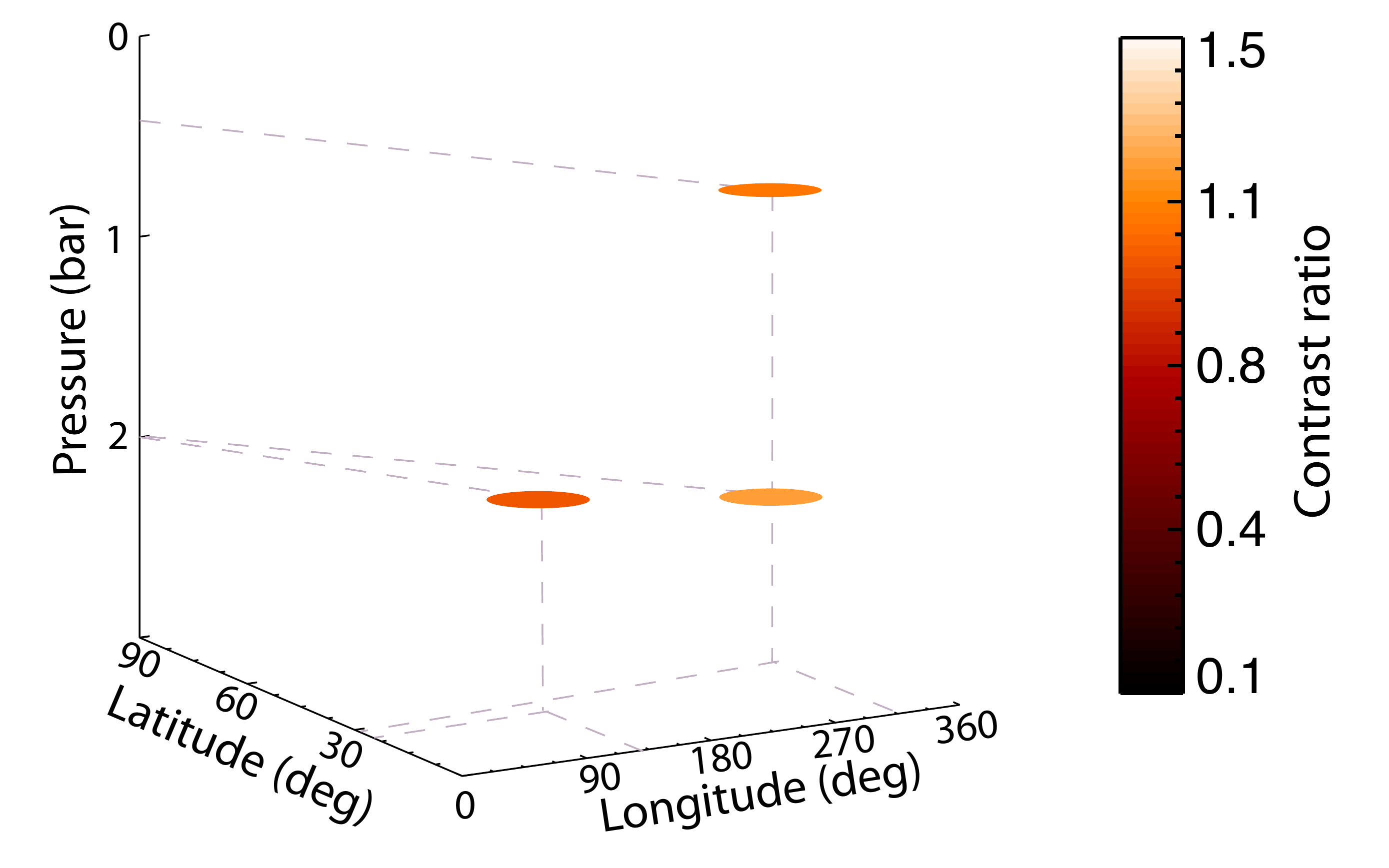}
\caption{Jupiter's model 3D structure based on our HST observations. 
From the observational light curves, \textit{Aeolus} retrieves a number of spots per 
wavelength/pressure level. Using contribution functions we define the pressure level 
from which most of the radiation comes from, and create a 3D map of the jovian atmosphere.}
\label{fig:jup_3d_struct}
\end{figure}

In this paper we presented \textit{Aeolus}, a Markov--Chain Monte Carlo code 
that maps the (2D) top--of--the--atmosphere (TOA) structure 
of brown dwarf and other directly detected ultra cool atmospheres, 
at a given observational wavelength. 
\textit{Aeolus} combines a Metropolis-Hastings algorithm with a Gibbs sampler and 
assumes that all heterogeneities at the TOA appear in the form of 
elliptical spots \citep[][]{cho96,cho08}. \textit{Aeolus} finds the number of spots needed to fit 
the observed light curve, and for each spot its size, contrast ratio to the 
background and location (latitude and longitude) on the disk.

We validated \textit{Aeolus} on the Jupiter dataset. \textit{Aeolus} retrieved accurately 
the major features observed in the Jovian atmosphere. 
\textit{Aeolus}, similarly to all flux--mapping techniques, 
cannot retrieve rotationally symmetric features (zones and belts of Jupiter) and 
suffers from latitudinal degeneracies \citep[see e.g.][]{apai13}. The latter is the 
reason why \textit{Aeolus} did not retrieve Oval BA (visible in the U--band), but found a slightly shifted 
latitude and larger size for the Great Red Spot (GRS). In the U--band \textit{Aeolus} 
retrieved the biggest, non--rotationally--symmetric 
feature of the jovian disk (in the U--band), the GRS. 
In the R--band \textit{Aeolus} retrieved the GRS and 
the largest 5 $\mu$m hot spot visible at the TOA. In both bands, smaller features such 
as high altitude NH$_3$ ice clouds, or smaller 5 $\mu$m hot spots were not retrieved. 
If we take into account that the Oval BA is large enough to influence the retrieved location and size 
of the GRS, then, the smallest feature retrieved by \textit{Aeolus} 
in our HST Jupiter dataset has a longitudinal extent of $\sim$11$^\circ$. 
\textit{Aeolus} is, to our knowledge, the first mapping code validated on actual 
observations of a giant planet over a full rotational period. 

Given the unprecedented high SNR (relative photometric per independent 
sample: $\sim$30,000) of these observations for the field of 
exoplanets and brown dwarfs, these results put constraints on the maximum size of TOA 
features we can map in the future using, for example, the James Webb Space Telescope (JWST). 
For example, modeling SIMP0136 spectrum as a black body (T$_\mathrm{eff}$=1100K), 
normalized so that M$_\mathrm{J}=$14.5 \citep[][]{artigau06,apai13} and assume 
we observe it with NIRCAM's F200W with a 12$\mathrm{s}$ exposure (resolution of 
0.5$^\circ$ of rotation for SIMP0136) we reach a SNR$\sim$4,160 (source: JWST 
prototype ETC, version P1.6). Considering that JWST will provide a finer cadence than HST 
can, the combined information content over a complete rotation on a high--amplitude variable 
brown dwarf will be comparable  with our current HST dataset on Jupiter. This suggests that mapping 
with an overall quality similar to that presented here may be possible for the most ideal brown dwarfs.
Assuming a contrast range in the atmospheric features that is similar to that observed in Jupiter in the 
visible, rotational maps could identify features $\sim$11$^\circ$ or larger, similar to the Oval BA in our study.  
\citep[][]{kostov13} argues that with high--contrast observations JWST will also be able to carry 
out analogous observations on directly imaged exoplanets.

We explored our Jupiter observations for the possibility that our temporal 
(i.e., rotational, or spatial) resolution affects the minimum size of the mapped 
features. In particular, we explored the possibility that the largest time gaps in 
our observations (corresponding to rotational ``jumps'' of 32.5$^\circ$ to 45.5$^\circ$ 
and are due to Earth occultations of the target (Jupiter) with HST's orbit) result in 
some of the features to not be followed throughout their rotation across the visible and 
illuminated disk, and to a lack of data during their appearance from, and/or disappearance to the dark 
side. Lack of ingress/egress data could influence the detectability of 
features, since the features' properties may not be well--constrained. We found that 
largest hot spot (that we detect) undergoes a similar ``jump'' of 32.5$^\circ$ to the dark side, 
implying that this should not be an important effect.

We then applied \textit{Aeolus} to two brown dwarfs in the L/T transition, 
2M2139 and SIMP0136. \citet[][]{apai13} obtained HST 
observations of these brown dwarfs and mapped them using principal component 
analysis (PCA) and the mapping package \textit{Stratos}. 
We compared \textit{Aeolus} results against \textit{Stratos} and Fourier mapping results. 
We found that, within the error bars, \textit{Aeolus} and \textit{Stratos} agree on the amount and  
coverage on the TOA of 2M2139 and SIMP0136, while Fourier mapping hints to larger 
longitudinal coverage.

A major difference between the \textit{Aeolus} and \textit{Stratos}+PCA results is that 
in the case of SIMP0136, \textit{Aeolus} retrieved a mix of brighter and darker than 
the background TOA spots, while \textit{Stratos}+PCA retrieved only one kind (brighter or darker) of spots. 
\textit{Aeolus} fits the properties of the spots on the TOA freely, and independently of each 
other, without any prior assumptions, while \textit{Stratos} uses PCA analysis to identify the 
smallest set of independent spectra (i.e., amount of different components/ surface contrast ratios), 
over the mean spectrum, that are needed to reproduce 96\% of 
the observed spectral variations. \citet[][]{apai13} using PCA found that only one spot--component 
is necessary to fit the observed variations of 2M2139 and SIMP0136, arguing that all spots 
are similar in nature \citep[see][]{apai13}. In contrast, \textit{Aeolus} found that the best-fit spots 
are composed of three (2M2139) or, potentially, two different surfaces (SIMP0136, taking into 
account the error bars). \citet[][]{apai13} using \textit{Stratos} 
found that there is a degeneracy between best-fit spot brightness and 
limb--darkening parameters and/or inclination of the brown dwarf. 
Given that in \textit{Aeolus} these parameters are fixed, the differences between 
the maps can be due to a wrong assumption for the inclination (we assumed $0^\circ$) 
or limb--darkening (we used $c\sim$0.5). 
In the future, we will upgrade \textit{Aeolus} to fit inclination and 
limb--darkening as free parameters. 

For a direct comparison with \textit{Stratos}, 
we ran a test case where we forced the contrast ratio of the spots to the background 
TOA to be the same for all spots. We kept the contrast ratio a factor 
of three brighter than the background TOA to match the \citet[][]{apai13} results. 
Our code retrieved 3 spots (BIC 52) covering 21.4$\pm$9.6\% of the top--of--the--atmosphere, 
in agreement with \textit{Stratos} results. The BIC for this solution is comparable 
to the best--fit model, making it an equally acceptable solution for \textit{Aeolus}. 
In the future, a synergy of \textit{Aeolus} with PCA can be used to control the validity 
of the best-fit models.

An interesting result is that, in agreement with the complexity of the light curves, 
both \textit{Aeolus} and \textit{Stratos} find that no 
one or two spot models can interpret the observed light curves accurately. This 
implies a complex TOA structure for both 2M2139 and SIMP0136. A similar, or 
more complex TOA structure was inferred for Luhman 16B \citep[][]{crossfield14}, 
and is also implied by the complex light curve shapes observed in other brown dwarfs 
\citep[see, e.g.,][]{metchev14}. This hints to complex dynamics in the atmospheres of brown 
dwarfs, which are predicted by models of atmospheric circulation \citep[][]{showman13,zhang14}.

As a demonstration of the potential for constraining atmospheric dynamics from rotational 
maps we briefly explore the possibility of constraining wind speeds from the maximum sizes 
of the features mapped, following a Rhines--length--based argument laid out in \citet[][]{apai13}, 
and also adopted in \citet[][]{burgasser14}. Our \textit{Aeolus}'  SIMP0136 and 2M2139 maps 
show features that are, on average, larger (in longitude/latitude) than the largest Jupiter feature. 
If we accept that our maps are accurate, and the retrieved spots uniform, this would imply a 
higher wind speed in the atmosphere of these brown dwarfs than in Jupiter's (assuming 
that the maximum spot size is set by the atmospheric jet widths). 
For example, using as the larger spot of Jupiter the GRS with s$_1\sim$17$^\circ$ 
and for 2M2139 s$_2\sim$39$^\circ$, Jupiter's period P$_1\sim$10 hrs and 2M2139's 
period P$_1\sim$7.61 hrs and the equatorial jet wind speed on Jupiter U$_1\sim$100 m/s, one can show 
that the wind speed on 2M2139 (assuming that the radius of 2M2139 is equal to Jupiter's radius) is 
$\sim$690 m/s. This speed is between the wind speeds of our Solar System giant planets [e.g., 
100 m/s for Jupiter, 500 m/s for Neptune \citet[][]{depaterlissauer10}] and highly irradiated exojupiters 
where wind speeds can reach a couple thousands of km/s \citep[e.g.,][]{snellen10, colon12}.
\citet[][]{radigan12} suggest a wind speed of 45 m/s for 
2M2139, even though they caution that their estimate may be offset and longer observation 
would be necessary to determine the actual wind speeds. For the slightly later 
T dwarf SIMP0136, \citet[][]{showman13} using \citet[][]{artigau09} input, find a wind speed 
of 300 m/s to 500 m/s. If these values are verified, they would suggest our largest mapped 
spots are ``blends'' of smaller spots, in a comparable way to \textit{Aeolus}' ``blend'' 
of Oval BA and the GRS.

An interesting result that emerged from the few brown dwarfs with high--quality simultaneous 
multi--wavelength observations is that light curves probing different pressure levels do not always 
line up with each other. Specifically, five light curves in  the late--T brown dwarf 2M2228 \citet[][]{buenzli12} 
observed between 1.1 and 5~$\mu$m showed a pressure--dependent phase lag. This was interpreted 
as evidence for large--scale longitudinal--vertical organization in the atmosphere \citet[][]{buenzli12}. 
Similar possible phase shift was reported in the L/T transition dwarf binary Luhman 16AB \citet[][]{biller13}. 
In contrast, the two L/T transition objects 2M2139 and SIMP0136 showed no phase shifts in the 
1.1--1.7~$\mu$m wavelength range, suggesting vertically identical surface brightness distribution 
\citep[][]{apai13}. Thus, the presence or absence of pressure--dependent phase shifts provides 
powerful constraints on the longitudal--vertical structure of the atmospheres.

Analogously, the different wavelength observations of the jovian atmosphere presented in our paper 
also probe different pressure levels. The differently shaped light curves reveal different surface 
brightness distributions (Fig.~\ref{fig:jup_3d_struct} and Sect.~\ref{sect:phenom_jup}). 
We note that if these light curves were observed with a SNR too low to allow distinguishing the 
differences in the light curve shape, the different peak times in the different light curves could be 
interpreted as phase shifts, even though they represent two uncorrelated structures. 

We propose that to ensure that uncorrelated light curves are not misinterpreted as phase shifts 
a crucial consideration should be the uncertainties along both the pressure and the phase shifts 
axes. The presence or absence of vertically organized atmospheric layers could be tested by 
comparing the goodness of a single trend versus multiple uncorrelated trends in the pressure--phase 
shift space, considering the error bars.

In this paper we presented two--dimensional maps of Jupiter and 
two brown dwarfs: 2M2139 and SIMP0136. \textit{Aeolus} though, can also 
produce three--dimensional maps of ultracool atmospheres. When the latter 
are observed at multiple wavelengths, \textit{Aeolus} can produce two--dimensional 
maps of the atmosphere per observational wavelength. Using information from a 
target--appropriate contribution function, we can identify the pressure level 
where most of the radiation emerges from [at that wavelength; see, e.g., \citet[][]{buenzli12}] and 
stack--up the two--dimensional maps. For example, in the case 
of Jupiter's HST observations, contribution functions suggest that the R--band 
originates around $2$ bars and the U--band around $400$ mbar. 
With this information and the \textit{Aeolus} retrieved maps, we can 
compose  a ``3D'' map of the modeled jovian atmosphere as in Fig.~\ref{fig:jup_3d_struct}. 
Studying the 3D structure of ultracool atmospheres and its variability over time is an important step 
towards understanding their dynamics. Long--scale atmospheric dynamical effects 
like cells and vortices, for example, will cause spots to move in 3D following the dynamical structure.
Using multiple--epoch, multi--wavelength observations and \textit{Aeolus} we can map 
the 3D structure of our targets over large periods and follow the 3D motions of structures 
in the atmospheres. These maps can then provide feedback to dynamical models, helping 
to study and understand dynamics governing ultracool atmospheres.

\textit{Aeolus} is a validated mapping code that can be used to map 
brown dwarf and directly imaged giant exoplanet atmospheres currently, and imaged 
terrestrial exoplanets in the future. For the latter, an adaptation of \textit{Aeolus} 
that takes into account surface (non--elliptical) structures would be necessary. 
Ideally, the updated version of \textit{Aeolus} would then be validated on a 
``ground truth'' dataset of Earth and/or Venus disk--integrated, multi--wavelength 
observations.

\textit{Aeolus} was, in part, developed to interpret observations from the 
Extrasolar Storms program (PI: Apai). Extrasolar Storms obtained multi--epoch HST and Spitzer 
observations of six brown dwarfs, to characterize cloud evolution and dynamics of brown 
dwarf atmospheres over multiple rotational periods. Extrasolar Storms observed six targets, 
in eight separate visits from Spitzer's IRAC channels 1 and 2, and two visits from HST 
WFC3 IR channel (G141). HST visits were coordinated with the Spitzer observations, 
so that for two visits we acquired multi--wavelength observations. We, currently, apply 
\textit{Aeolus} on the full Extrasolar Storms sample and will publish our results in a 
follow--up paper.

\section{Conclusions}

We presented \textit{Aeolus}, a Markov--Chain Monte Carlo code 
that maps the two--dimensional top--of--the--atmosphere structure 
of brown dwarf and other directly detected ultra cool atmospheres, 
at a given observational wavelength. We validated \textit{Aeolus} on 
a unique spatially and temporally resolved imaging data 
set of the full disk of Jupiter in two spectral bands. This data set provides a``truth test'' to 
validate mapping of ultracool atmospheres by \textit{Aeolus} and any other 
mapping methods/ tools. The dataset will be publicly available via ADS/VIZIR. 
\textit{Aeolus} is the first mapping code validated on actual observations 
of a giant planet over a full rotational period.

We noted that if our Jupiter light curves were observed with a signal--to--noise--ratio 
too low to allow distinguishing the differences in the light curve shape, the different peak 
times in the different light curves could be interpreted as phase shifts, analogous to the ones 
seen in 2M2228, even though they represent two uncorrelated structures. 
To ensure that uncorrelated light curves are not misinterpreted 
as phase shifts we need better constrains of the uncertainties along both the 
pressure and the phase shifts axes. 

Finally, we applied \textit{Aeolus} to 2M2139 and SIMP0136. \textit{Aeolus} found 
three spots at the top--of--the--atmosphere of these two brown dwarfs, 
with a coverage of 21\%$\pm$3\% and 20.3\%$\pm$1.5\% respectively, 
in agreement with previous mapping efforts. 
Constraining wind speeds from the maximum sizes of the features
in \textit{Aeolus}' maps we retrieved a wind speed of $\sim$690 m/s for 2M2139. 
Observations of 2M2139 and SIMP0136 suggest lower wind speeds, up to 500 m/s, 
which, if confirmed, imply that \textit{Aeolus}' largest features mapped are blends of 
smaller spots.

 \acknowledgements
This work is part of the Spitzer Cycle-9 Exploration Program
Extrasolar Storms (program No. 90063). This work is based on observations made with the Spitzer Space 
Telescope, which is operated by the Jet Propulsion Laboratory, California Institute of 
Technology under a contract with NASA. Support for this work was provided by 
NASA through an award issued by JPL/Caltech.
Support for Program number 12314 was provided by NASA
through a grant from the Space Telescope Science Institute,
which is operated by the Association of Universities for
Research in Astronomy, Incorporated, under NASA contract
NAS5-26555. 
An allocation of computer time from the UA Research Computing High Performance 
Computing (HTC) and High Throughput Computing (HTC) at the University of Arizona 
is gratefully acknowledged. This study, in part, is based on observations 
made with the NASA/ ESA Hubble Space Telescope, obtained at the Space 
Telescope Science Institute, which is operated by the Association of Universities 
for Research in Astronomy, Inc., under NASA contract NAS 5Ð26555. These 
observations are associated with program No. 13067. GS and JMP acknowledge 
support for program No. 13067 provided by NASA through grants from 
the Space Telescope Science Institute to the University Arizona and to 
Williams College, respectively. JMP acknowledges the hospitality of A. Ingersoll and the Planetary Sciences Department of Caltech.



\end{document}